\documentclass[fleqn,usenatbib]{mnras}
\usepackage{newtxtext,newtxmath}
\usepackage[T1]{fontenc}
\usepackage{ae,aecompl}

\usepackage{graphicx}
\graphicspath{{Figures/}}
\usepackage{amsmath}	
\usepackage{amssymb}
\usepackage{wasysym}
\usepackage{color}
\usepackage{tabularx}
\newcolumntype{C}{>{\centering\arraybackslash}X}
\newcolumntype{P}[1]{>{\centering\arraybackslash}p{#1}}
\defcitealias{Schaye2015}{S15}

\title[C-EAGLE ICM properties]{The Cluster-EAGLE project: global properties of simulated clusters with resolved galaxies}

\author[Barnes et al.]{David J. Barnes$^{1}$\thanks{E-mail: david.barnes@manchester.ac.uk}, Scott T. Kay$^{1}$, Yannick M. Bah\'e$^{2}$, Claudio Dalla Vecchia$^{3,4}$,\newauthor Ian G. McCarthy$^{5}$, Joop Schaye$^{6}$, Richard G. Bower$^{7}$, Adrian Jenkins$^{7}$, \newauthor Peter A. Thomas$^{8}$, Matthieu Schaller$^{7}$, Robert A. Crain$^{5}$, Tom Theuns$^{7}$ \newauthor and Simon D. M. White$^{2}$
\\
$^{1}$Jodrell Bank Centre for Astrophysics, School of Physics and Astronomy, The University of Manchester, Manchester M13 9PL, UK\\
$^{2}$Max-Planck-Institut f\"ur Astrophysik, Karl-Schwarzschild Str. 1, 85748 Garching, Germany \\
$^{3}$Instituto de Astrof\'\i{}sica de Canarias, C/ V\'\i{}a L\'actea s/n, 38205 La Laguna, Tenerife, Spain\\
$^{4}$Departamento de Astrof\'\i{}sica, Universidad de La Laguna, Av.~del Astrof\'\i{}sico Francisco S\'anchez s/n, 38206 La Laguna,
Tenerife, Spain\\
$^{5}$Astrophysics Research Institute, Liverpool John Moores University, 146 Brownlow Hill, Liverpool, L3 5RF, UK \\
$^{6}$Leiden Observatory, Leiden University, PO Box 9513, 2300 RA Leiden, The Netherlands \\
$^{7}$Institute for Computational Cosmology, Department of Physics, University of Durham, South Road, Durham DH1 3LE, UK \\
$^{8}$Astronomy Centre, University of Sussex, Falmer, Brighton BN1 9QH, UK \\
}

\date{Accepted XXX. Received YYY; in original form ZZZ}

\pubyear{2017}

\begin{document}
\label{firstpage}
\pagerange{\pageref{firstpage}--\pageref{lastpage}}
\maketitle

\begin{abstract}
We introduce the Cluster-EAGLE (\textsc{c-eagle}) simulation project, a set of cosmological hydrodynamical zoom simulations of the formation of $30$ galaxy clusters in the mass range $10^{14}<M_{200}/\mathrm{M}_{\astrosun}<10^{15.4}$ that incorporates the {\it Hydrangea} sample of Bah\'e et al. (2017). The simulations adopt the state-of-the-art \textsc{eagle} galaxy formation model, with a gas particle mass of $1.8\times10^{6}\,\mathrm{M}_{\astrosun}$ and physical softening length of $0.7\,\mathrm{kpc}$. In this paper, we introduce the sample and present the low-redshift global properties of the clusters. We calculate the X-ray properties in a manner consistent with observational techniques, demonstrating the bias and scatter introduced by using estimated masses. We find the total stellar content and black hole masses of the clusters to be in good agreement with the observed relations. However, the clusters are too gas rich, suggesting that the AGN feedback model is not efficient enough at expelling gas from the high-redshift progenitors of the clusters. The X-ray properties, such as the spectroscopic temperature and the soft-band luminosity, and the Sunyaev-Zel'dovich properties are in reasonable agreement with the observed relations. However, the clusters have too high central temperatures and larger-than-observed entropy cores, which is likely driven by the AGN feedback after the cluster core has formed. The total metal content and its distribution throughout the ICM are a good match to the observations.
\end{abstract}


\section{Introduction}
\label{sec:intro}
The observable properties of galaxy clusters emerge from the complex interplay of astrophysical processes and gravity acting on hierarchically increasing scales. Cluster formation is a process that has an enormous dynamic range, as clusters collapse from fluctuations with a co-moving scale length of tens of Mpc, but have observable properties that are shaped by highly energetic astrophysical processes acting on sub-parsec scales \citep[see][for recent reviews]{Voit2005,AllenEvrardMantz2011,KravtsovBorgani2012}. The interaction of processes acting on very different scales makes cluster formation a highly non-linear process. However, the combination of scales and processes make galaxy clusters a unique environment where we can observe not only the material that participated in galaxy formation, but also the material that did not. Therefore, clusters allow the simultaneous study of fundamental cosmological parameters, gravity, hydrodynamical effects, chemical element synthesis, and the interaction of relativistic jets with the cluster environment.

Although significant progress can be made with semi-analytic prescriptions \citep{Bower2008,Somerville2008,Guo2011,Bower2012}, hydrodynamical simulations are the only method that can capture the effects of physical processes during cluster formation and predict the resulting observable consequences self-consistently. Although unable to capture the full dynamic range due to limited computational resources, there has been significant progress in the modelling of cluster formation and the physical processes that occur below the resolution scale of the simulation, so called subgrid models. The formation of the baryonic component of clusters has been well studied \citep[e.g.][]{EkeNavarroFrenk1998,Kay2004,NagaiVikhlininKravtsov2007a,Crain2007,Sijacki2007,Short2010,Dubois2010,Young2011,Battaglia2012,Gupta2016}, including the importance of feedback from active galactic nuclei (AGN) and its effect on the baryonic content of clusters \citep[e.g.][]{PuchweinSijackiSpringel2008,Fabjan2010,McCarthy2010,Martizzi2016}. These developments have led to several independent groups simulating samples of clusters that are to varying degrees realistic \citep{LeBrun2014,Pike2014,Planelles2014,Rasia2015,Hahn2017}, i.e. their observable properties, such as X-ray luminosity and spectroscopic temperature, are a good match to those of observed clusters. The understanding that subgrid models should be calibrated against carefully selected observable relations has resulted in simulations that simultaneously reproduce a host of stellar, gas and halo properties \citep{McCarthy2017}, even to high redshift \citep{Barnes2017}. One limitation of previous cluster formation simulation work is that it only achieved a modest resolution, typically with a gas particle mass of $m_{\mathrm{gas}}\sim10^{9}\,\mathrm{M}_{\astrosun}$ (for smoothed particle hydrodynamic, SPH, simulations) and a spatial resolution of $\sim5\,\mathrm{kpc}$. This limits the ability to resolve the structures in the intracluster medium (ICM), to examine the interactions between energetic astrophysical processes and the ICM, to capture the formation and evolution of the cluster galaxy population, and to resolve the growth histories of the black holes.

At the same time there has been significant progress in the theoretical modelling of galaxy formation in representative volumes. Improved resolution, and the development and calibration of efficient subgrid prescriptions for feedback processes have led to a step change in the realism of galaxy formation models \citep{Vogelsberger2014,Schaye2015,Crain2015,Dave2016,Tremmel2016}. For example, the \textsc{eagle} simulation suite \citep{Schaye2015,Crain2015} was calibrated against the observed galaxy stellar mass function, the field galaxy size-mass relation and the black hole mass-stellar mass relation at low redshift. Following this, the model then yields broad agreement with, among other things, the observed evolution of galaxy star formation rates \citep{Furlong2015}, the evolution of galaxy sizes \citep{Furlong2017}, their molecular and atomic hydrogen content \citep{Lagos2015,Bahe2016,Crain2017}, their observed colour distribution \citep{Trayford2015}, and the growth of black holes and their link to the star formation and growth of galaxies \citep{RosasGuevara2016,Bower2017,McAlpine2017}. However, the resolution required and complexity of the subgrid models makes these simulations computationally expensive, limiting their volume to periodic cubes with a side length of $\sim100\,\mathrm{Mpc}$ or less. Although a volume of this size will contain many galaxy groups ($M_{200}=10^{13}-10^{14}\,\mathrm{M}_{\astrosun}$\footnote{We define $M_{200}$ as the mass enclosed within a sphere of radius $r_{200}$ whose mean density is $200$ times the critical density of the Universe.}), rich galaxy clusters $(M_{200}\geq10^{15}\,\mathrm{M}_{\astrosun})$ are very rare objects and a volume of this size is highly unlikely to contain even one. Therefore, it is difficult to assess the ability of these calibrated models to produce realistic large-scale structures, such as galaxy clusters, and to test whether they correctly capture galaxy formation in the full range of environments.

Motivated by the limitations of existing cluster and galaxy formation simulations, we introduce the Virgo consortium's Cluster-EAGLE (\textsc{c-eagle}) project. The project consists of zoom simulations of the formation of $30$ galaxy clusters that are evenly spaced in the mass range $10^{14}-10^{15.4}\,\mathrm{M}_{\astrosun}$, probing environments that are not present in the original periodic \textsc{eagle} volumes presented by \citet{Schaye2015}, henceforth \citetalias{Schaye2015}, and \citet{Crain2015}. They are performed with the \textsc{eagle} galaxy formation model (AGNdT9 calibration) and adopt the same mass resolution ($m_{\mathrm{gas}}=1.81\times10^{6}\,\mathrm{M}_{\astrosun}$) and physical spatial resolution ($\epsilon=0.7\,\mathrm{kpc}$) as the largest periodic volume of the \textsc{eagle} suite (Ref-L100N1504). The resolution of the simulations allows us to resolve the formation of cluster galaxies and their co-evolution with the ICM, the interactions between the cluster galaxies and the ICM, the formation of structures within the ICM, and how energetic astrophysical processes shape the ICM. As the hot halo typically extends to several virial radii the zoom regions extend to at least five times the virial radius of each object to include the large-scale structure around them. The {\it Hydrangea} sample \citep{Bahe2017}, designed to study the evolution of galaxies as their environment transitions from isolated field to dense cluster, extends the zoom region to ten virial radii for $24$ of $30$ \textsc{c-eagle} clusters.

In this paper, we present the global properties and hot gas profiles of the simulated clusters at low-redshift and compare to observations in order to examine the ability of a model calibrated for galaxy formation to produce realistic galaxy clusters. In a companion paper \citep{Bahe2017} the properties of the cluster galaxy population are presented and the {\it Hydrangea} sample is used to study the impact of the cluster environment on the galaxy stellar mass function. The predicted galaxy luminosity functions of the clusters will be presented in Dalla Vecchia et al. ({\it in prep.}), including results for higher-resolution runs of a subset of the clusters. The rest of this paper is structured as follows. In Section \ref{sec:NumMeth} we present the sample selection, a brief overview of the \textsc{eagle} model and the method adopted for computing global properties in a manner consistent with observational techniques, which enables a fairer comparison to observational data. We then compare the global properties of the sample to observational data in Section \ref{sec:screlations} and examine the hot gas profiles of the sample in Section \ref{sec:gasprofs}. Finally, we discuss our results in Section \ref{sec:Dis} and present a summary of the main findings in Section \ref{sec:Sum}.

\section{Numerical Method}
\label{sec:NumMeth}
This section provides an overview of the cluster sample selection, the model used to resimulate them and how the observable properties were calculated in a manner consistent with observational approaches.

\subsection{Sample selection}
Due to their limited size the original \textsc{eagle} volumes contain very few clusters, with the largest volume (Ref-L100N1504) containing $7$ objects with a mass $M_{200}>10^{14}\,\mathrm{M}_{\astrosun}$. The \textsc{c-eagle} project provides an extension to the cluster environment by performing zoom simulations, which requires a population of clusters that a representative sample can be selected from. We use the parent simulation from \citet{Barnes2017} as the basis of our sample selection. It uses a \textit{Planck} 2013 cosmology \citep{Planck2014I} and is a cubic periodic volume with a side length of $3.2\,\mathrm{Gpc}$, which is large enough to contain the rarest and most massive haloes expected to form in a $\Lambda\mathrm{CDM}$ cosmology. At $z=0$ it contains $185,150$ haloes with $M_{200}>10^{14}\,\mathrm{M}_{\astrosun}$ and $1701$ haloes with $M_{200}>10^{15}\,\mathrm{M}_{\astrosun}$. The sample was selected by first binning all haloes into ten evenly spaced log mass bins in the range $14.0\leq\log_{10}(M_{200}/\mathrm{M}_{\astrosun})\leq15.4$. We did this to ensure that we evenly sampled the chosen mass range, otherwise we would have been biased towards lower masses by the steep slope of the mass function. To ensure that our selected objects would be at the centre of the peak in the local density structure and the focus of our computational resources, we removed objects from the selection bins who had a more massive neighbour within a sphere whose radius was the larger value of either $30\,\mathrm{Mpc}$ or $20\,r_{200}$. We then randomly picked three haloes from each mass bin to yield a sample of $30$ objects, which are listed in Appendix \ref{app:table}.

We used the zoom simulation technique \citep{KatzWhite1993,Tormen1997} to resimulate our chosen sample at higher resolution. The Lagrangian region of every cluster was selected so that its volume was devoid of lower resolution particles beyond a cluster-centric radius of at least $5\,r_{200}$ at $z=0$. Additionally, the Lagrangian regions of the {\it Hydrangea} sample were defined such that they were devoid of lower resolution particles beyond a cluster-centric radius of $10\,r_{200}$, enabling studies of galaxy evolution as the environment transitions from isolated field to dense cluster. At $z=127$, the initial glass-like particle configuration of the high-resolution regions was deformed according to second-order Lagrangian perturbation theory using the method of \citet{Jenkins2010} and \textsc{Panphasia} \citep{Jenkins2013}, a multi-scale Gaussian white noise field that is publicly available\footnote{The phase descriptor of the parent volume is given in Appendix \ref{app:table} and in Table B1 of \citetalias{Schaye2015}.}. We assumed a flat $\Lambda\mathrm{CDM}$ cosmology based on the \textit{Planck} 2013 results combined with baryonic acoustic oscillations, WMAP polarization and high multipole moments experiments \citep{Planck2014I}. The cosmological parameters were $\Omega_{\rm{b}}=0.04825$, $\Omega_{\rm{m}}=0.307$, $\Omega_{\Lambda}=0.693$, $h\equiv H_0/(100\,\rm{km}\,\rm{s}^{-1}\,\rm{Mpc}^{-1})=0.6777$, $\sigma_{8}=0.8288$, $n_{\rm{s}}=0.9611$ and $Y=0.248$. The resolution of the Lagrangian regions was increased to match the resolution of the \textsc{eagle} $100\,\mathrm{Mpc}$ simulation (Ref-L100N1504). The dark matter particles each had a mass of $m_{\mathrm{DM}}=9.7\times10^{6}\,\mathrm{M}_{\astrosun}$ and the gas particles each had an initial mass of $m_{\mathrm{gas}}=1.8\times10^{6}\,\mathrm{M}_{\astrosun}$ (note no $h^{-1}$). The proper gravitational softening length for the high-resolution region was set to $2.66$ comoving $\mathrm{kpc}$ for $z>2.8$, and then kept fixed at $0.70$ physical $\mathrm{kpc}$ for $z<2.8$. The minimum smoothing length of the SPH kernel was set to a tenth of the gravitational softening scale.

\subsection{The EAGLE model}
We use the \textsc{eagle} model to resimulate our selected sample. The \textsc{eagle} subgrid model is based on the model developed for the \textsc{owls} \citep{Schaye2010} project and also used for the \textsc{gimic} \citep{Crain2009} and \textsc{cosmo-owls} \citep{LeBrun2014} models. The subgrid model, the calibration of its free parameters and its numerical convergence are described in detail in \citetalias{Schaye2015} and \citet{Crain2015}. The code is a heavily modified version of the \textit{N}-body Tree-PM SPH code \textsc{P-Gadget-3}, which was last described in \citet{Springel2005}. The hydrodynamics algorithms are collectively known as `\textsc{anarchy}' (see Dalla Vecchia (in prep.), appendix A of \citetalias{Schaye2015} and \citet{Schaller2015}) and consists of an implementation of the pressure-entropy SPH formalism derived by \citet{Hopkins2013}, an artificial viscosity switch \citep{CullenDehnen2010}, an artificial conductivity switch similar to that of \citet{Price2008}, the $C^2$ smoothing kernel with $58$ neighbours \citep{Wendland1995} and the time-step limiter of \citet{DurierDallaVecchia2012}. The subgrid model includes radiative cooling, star formation, stellar evolution, feedback due to stellar winds and supernovae, and the seeding, growth and feedback from black holes. We now briefly describe the subgrid model in more detail.

Net cooling rates are calculated on an element-by-element basis following \citet{WiersmaSchayeSmith2009}, under the assumption of an optically thin gas in ionization equilibrium, the presence of the cosmic microwave background and an evolving ultra-violet/X-ray background \citep{HaardtMadau2001} from galaxies and quasars. This is done by interpolation tables, computed using \textsc{cloudy} version 07.02 \citep{Ferland1998}, that are a function of density, temperature and redshift for the $11$ elements that were found to be important. During reionization, $2\,\mathrm{eV}$ per proton mass is injected to account for enhanced photo-heating rates. For hydrogen this occurs instantaneously at $z=11.5$ and for helium this additional heating is Gaussian distributed in redshift, centred on $z=3.5$ with a width $\sigma(z)=0.5$. The later ensures that the observed thermal history of the intergalactic gas is broadly reproduced \citep{Schaye2000,Wiersma2009}.

Star formation is modelled stochastically in a way that, by construction, reproduces the observed Kennicutt-Schmidt relation, as cosmological simulations lack the resolution and physics to properly model the  cold interstellar gas phase. It is implemented as a pressure-law \citep{SchayeDallaVecchia2008}, subject to a metallicity-dependent density threshold \citep{Schaye2004}. Gas particles whose density exceeds $n_{\mathrm{H}}(Z) = 10^{-1}\mathrm{cm}^{-3}(Z/0.002)^{-0.64}$, where $Z$ is the gas metallicity, are eligible to form stars. Lacking the resolution and physics to model the cold gas phase, a temperature floor, $T_{\mathrm{eos}}(\rho_{\mathrm{g}})$, is imposed that corresponds to the equation of state $P_{\mathrm{eos}}\propto\rho^{4/3}_{\mathrm{g}}$, normalized to $T_{\mathrm{eos}}=8\times10^{3}\,\mathrm{K}$ at $n_{\mathrm{H}}=10^{-1}\mathrm{cm}^{-3}$. This helps to prevent spurious fragmentation. Stellar evolution and the resulting chemical enrichment is based upon \citet{Wiersma2009}. Star particles are treated as simple stellar populations with a mass range of $0.1-100\,\mathrm{M}_{\astrosun}$ and a \citet{Chabrier2003} initial mass function. Mass loss and the release of $11$ chemical elements due to winds from massive stars, asymptotic giant branch stars and type Ia and type II supernovae are tracked. Feedback from star formation is implemented using the stochastic thermal model of \citet{DallaVecchiaSchaye2012}. The energy injected heats a particle by a fixed temperature increment, $\Delta T=10^{7.5}\,\mathrm{K}$, to prevent spurious numerical losses. The energy per unit of stellar mass formed, which sets the probability of heating events, depends on the local gas density and metallicity, and is calibrated to ensure that the galaxy stellar mass function and galaxy size-mass relation are a good match to the observed relations at $z=0.1$ \citep{Crain2015}.

Feedback from supermassive black holes (BHs) is a critical component of structure formation simulations, shaping the bright end of the galaxy luminosity function \citep[e.g.][]{Croton2006,Bower2006}, the gas content of clusters \citep[e.g.][]{PuchweinSijackiSpringel2008,McCarthy2010,Fabjan2010} and preventing the onset of the overcooling problem \citep{McCarthy2011}. The prescription for the seeding, growth and feedback from BHs is based on \citet{SpringelDiMatteoHernquist2005} with modifications from \citet{BoothSchaye2009} and \citet{RosasGuevara2015}. Seed BHs are placed in the centre of every halo with a total mass greater than $10^{10}\,\mathrm{M}_{\astrosun}/h$, with the highest density gas particle being converted to a BH particle with a subgrid seed mass of $m_{\mathrm{BH}}=10^{5}\,\mathrm{M}_{\astrosun}/h$. Haloes are identified by an on-the-fly friends-of-friends algorithm with a linking length of $b=0.2$ \citep{Davis1985}. BHs can grow either by merging with other BHs or via the accretion of gas. The gas accretion rate is Eddington limited and depends on the mass of the BH, the local gas density and temperature, and the relative velocity and angular momentum of the gas compared to the BH
\begin{equation}
 \dot{m}_{\mathrm{accr}}=\dot{m}_{\mathrm{Bondi}}\times\mathrm{min}(C^{-1}_{\mathrm{visc}}(c_{\mathrm{s}}/V_{\phi})^{3},1)\:,
\end{equation}
where $\dot{m}_{\mathrm{Bondi}}$ is the \citet{BondiHoyle1944} spherically symmetric accretion rate, $c_{\mathrm{s}}$ is the sound speed of the gas and $V_{\phi}$ is the rotation speed of the gas around the BH \citep[see equation 16 of][]{RosasGuevara2015}. $C_{\mathrm{visc}}$ is a free parameter for the effective viscosity of the subgrid accretion disc/torus and larger values correspond to a lower kinetic viscosity, which delays the growth of BHs and the onset of feedback events. We note that the results are remarkably insensitive to a non-zero value of $C_{\mathrm{visc}}$ \citep{Bower2017}. The growth of the BH is then given by $\dot{m}_{\mathrm{BH}}=(1-\epsilon_{\mathrm{r}})\dot{m}_{\mathrm{accr}}$, where $\epsilon_{\mathrm{r}}=0.1$ is the radiative efficiency. The accretion rate is not multiplied by a constant or density dependent factor as is common to many previous studies \citep{SpringelDiMatteoHernquist2005,BoothSchaye2009} as at the resolution of \textsc{eagle} we resolve sufficiently high gas densities and the accretion rate is sufficient for the BH seeds to grow by Bondi-Hoyle accretion.

\renewcommand\arraystretch{1.5}
\begin{table}
 \centering
 \caption{Values of the AGN feedback free parameters of the AGNdT9 model, used for the \textsc{\textsc{c-eagle}} simulations and the $50\,\mathrm{Mpc}$ \textsc{eagle} volume, and the REF model, used for the $100\,\mathrm{Mpc}$ \textsc{eagle} volume.}
 \begin{tabular}{l c c c}
  \hline
  Model & $n_{\mathrm{heat}}$ & $C_{\mathrm{visc}}$ & $\Delta T$ $[\mathrm{K}]$ \\
  \hline
  AGNdT9 & 1 & $2\mathrm{\pi}\times10^{2}$ & $10^{9}$ \\
  REF & 1 & $2\mathrm{\pi}$ & $10^{8.5}$ \\
  \hline
 \end{tabular}
 \label{tab:AGNdT9}
\end{table}
\renewcommand\arraystretch{1.0}

AGN feedback is implemented in a similar way to stellar feedback, whereby thermal energy is injected stochastically. Energy is injected at a rate of $\epsilon_{\mathrm{f}}\epsilon_{\mathrm{r}}\dot{m}_{\mathrm{accr}}c^{2}$, where $c$ is the speed of light and $\epsilon_{\mathrm{f}}=0.15$ is the fraction of energy that couples to the gas. \citet{BoothSchaye2009} found that for \textsc{owls} this value yielded agreement with the observed BH masses and \citetalias{Schaye2015} found that the same holds for \textsc{eagle}. In order to prevent spurious numerical losses and enable the feedback to do work on the gas before the injected energy is radiated, the BH stores accretion energy until it reaches a critical energy, at which point it has enough energy to heat $n_{\mathrm{heat}}$ particles by a temperature $\Delta T$. In the \textsc{eagle} model $n_{\mathrm{heat}}=1$. In the pressure-entropy SPH formalism used in \textsc{eagle}, the weighted density and entropic function are coupled \citepalias[][see appendix A1.1]{Schaye2015}. For large changes in internal energy, such as AGN feedback events, the iterative scheme used to adjust the density and entropic function fails to adequately conserve energy if many particles are heated simultaneously. This leads to a violation of energy conservation. The probability of heating a neighbour particle is given by
\begin{equation}
 P=\frac{E_{\mathrm{BH}}}{\Delta\epsilon_{\mathrm{AGN}}N_{\mathrm{ngb}}\langle m_{\mathrm{gas}}\rangle}\,,
\end{equation}
where $E_{\mathrm{BH}}$ is the energy reservoir of the BH, $\Delta\epsilon_{\mathrm{AGN}}$ is the specific energy associated with heating a particle by $\Delta T$, $N_{\mathrm{ngb}}$ is the number of gas neighbours and $\langle m_{\mathrm{gas}}\rangle$ is their average mass. As $E_{\mathrm{BH}}$ increases the probability of heating many neighbours increases and a violation of energy conservation becomes more likely. To prevent this the probability is capped at a value of $P_{\mathrm{AGN}}=0.3$, a value that was chosen following testing of the iterative scheme. If the unused energy remains above the critical threshold for a feedback event then the time step of the BH is shortened and the energy spread out over successive steps.

\begin{figure*}
 \includegraphics[width=\textwidth,keepaspectratio=True]{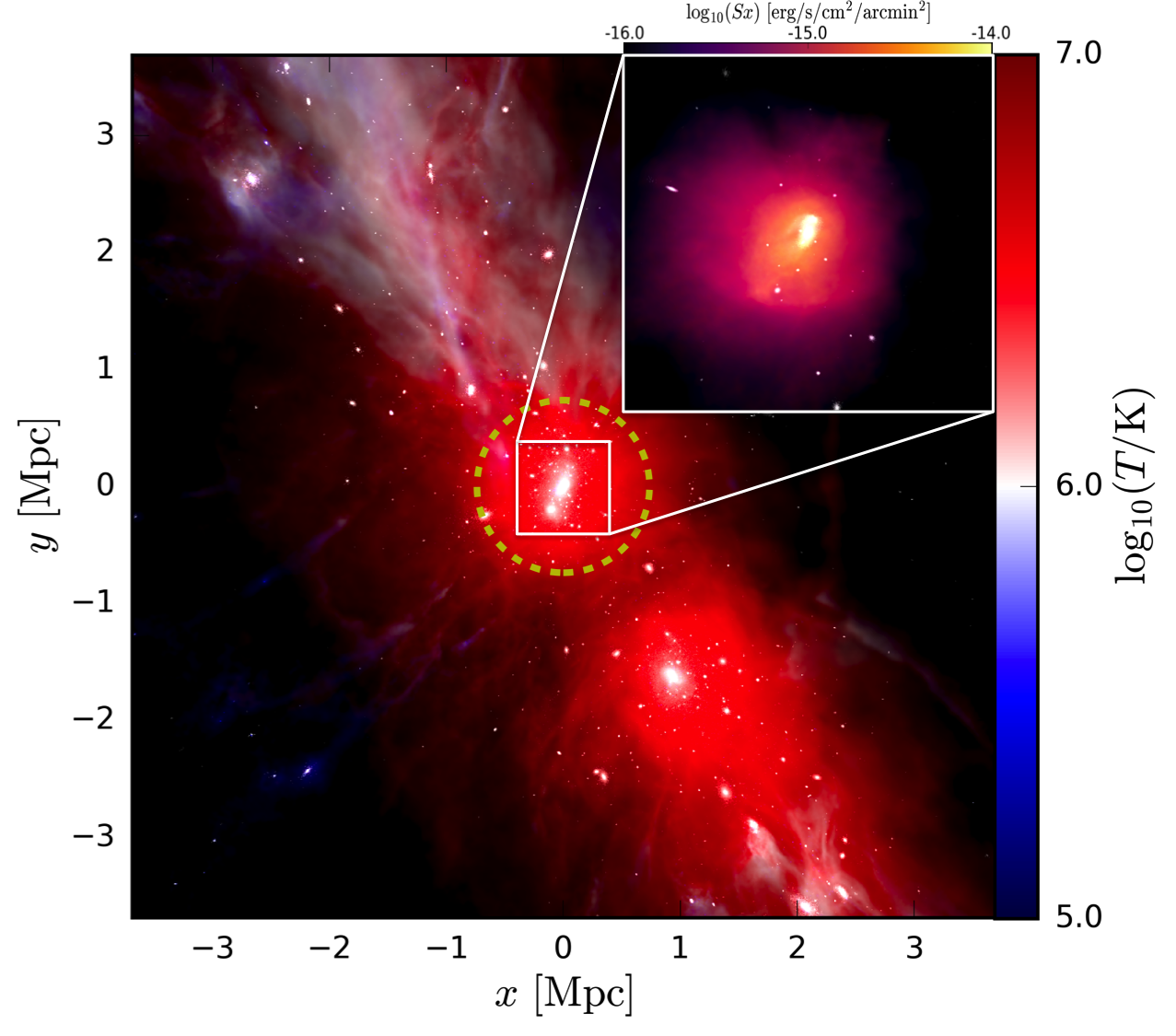}
 \caption{Image of CE-$5$ and its environment at $z=0.1$, resimulated using the \textsc{eagle} AGNdT9 model. The colour map shows the gas, with the intensity depicting the density of the gas and the colour depicting the temperature of the gas. Stellar particles are shown by the white points and the dashed yellow circle denotes $r_{200}$. The inset colour map shows the X-ray surface brightness from a cubic region of $2\,r_{500}$, centred on the cluster's centre of potential.}
 \label{fig:cluster_image}
\end{figure*}

\citetalias{Schaye2015} present three calibrated models (REF, AGNdT9 and Recal) that produce a similarly good match to the observed galaxy stellar mass function and galaxy mass-size relation. As we are running at standard \textsc{eagle} resolution we ignore the Recal model, which is relevant for simulations with $8\times$ higher mass resolution. From figures $15$ and $16$ of \citetalias{Schaye2015} it is clear that the AGNdT9 model provides a better match to the observed gas fraction-total mass and X-ray luminosity-temperature relations of low-mass groups ($M_{500}<10^{13.5}\,\mathrm{M}_{\astrosun}$). Therefore, we select the AGNdT9 model as our fiducial model for the \textsc{c-eagle} project; the free parameters of the AGN feedback for this model and for the REF model are given in Table \ref{tab:AGNdT9}. This model was calibrated in a $50\,\mathrm{Mpc}$ cubic volume to produce good agreement between simulated and observed galaxies. Furthermore, \citet{Crain2015} showed that the simulated BH-stellar mass relation and galaxy mass-size relation are sensitive to the feedback parameters used. We do not recalibrate the model to cluster scale objects and instead choose to retain the match to the observed galaxy properties. Therefore, the properties of the intracluster medium for the \textsc{c-eagle} clusters are a prediction of a model that produces reasonably realistic field galaxies.

\subsection{Calculating X-ray and SZ properties}
Inhomogeneities in the hot gas, e.g. the presence of multi-temperature structures, can significantly bias the ICM properties inferred from X-ray observations \citep[e.g.][]{NagaiVikhlininKravtsov2007b,Khedekar2013}. Therefore, when comparing simulations to observations it is vital to make a like-with-like comparison and compute properties from simulated clusters in a manner more consistent with observational techniques.

Following \citet{LeBrun2014}, we produce mock X-ray spectra for each cluster by first computing a rest-frame X-ray spectrum in the $0.05-100.0\,\mathrm{keV}$ band for each gas particle, using their individual density, temperature and SPH smoothed metallicity. We use the Astrophysical Plasma Emission Code \citep[\textsc{apec};][]{Smith2001} via the \textsc{pyatomdb} module with atomic data from \textsc{atomdb} v3.0.3 \citep[last described in][]{Foster2012}. For each of the $11$ elements we calculate an individual spectrum, ignoring particles with a temperature less than $10^5\,\mathrm{K}$ as they will have negligible X-ray emission.

Particles are then binned, in $3\mathrm{D}$, into $25$ logarithmically spaced radial bins centred on the potential minimum and the spectra are summed for each bin. The binned spectra are scaled by the relative abundance of the heavy elements as the fiducial spectra assume solar abundances specified by \citet{AndersGrevesse1989}. The energy resolution of a spectrum is $150\,\mathrm{eV}$ between $0.05-10.0\,\mathrm{keV}$ and we use a further $10$ logarithmically spaced bins between $10.0-100.0\,\rm{keV}$. A single temperature, fixed metallicity \textsc{apec} model is then fitted to the spectrum in the range $0.5-10.0\,\mathrm{keV}$, for each radial bin, to derive an estimate of the density, temperature and metallicity. During the fit we multiply the spectrum by the effective area of \textit{Chandra} for each energy bin to provide a closer match to typical X-ray observations.

We then perform a hydrostatic analysis of each cluster using the X-ray derived density and temperature profiles. We fit the density and temperature models proposed by \citet{Vikhlinin2006} to obtain a hydrostatic mass profile. We then estimate various cluster masses and radii, such as $M_{500}$ and $r_{500}$, from the hydrostatic analysis. We calculate properties, such as gas mass, $M_{\mathrm{gas},500}$, or Sunyaev-Zel'dovich flux, $Y_{\mathrm{SZ},500}$, by summing the properties of the particles that fall within the estimated X-ray apertures. A core-excised quantity is calculated by summing particles that fall in the radial range $0.15-1.0$ of the specified aperture. To calculate X-ray luminosities we integrate the spectra of particles that fall within the aperture in the required energy band, for example the soft band luminosities are calculated between $0.5-2.0\,\rm{keV}$. To estimate a spectroscopic X-ray temperature within an aperture, we sum the spectra of all particles that fall within it and then fit a single temperature \textsc{apec} model to the combined spectrum. All quantities that are derived in this way are labeled as `spec' quantities. In Appendix \ref{app:table} we also provide the values of estimated quantities for each cluster at $z=0.1$.

In Fig. \ref{fig:cluster_image} we show an image of the gas and stars for CE-$5$ at $z=0.1$. The resolution and subgrid physics of the \textsc{eagle} model enable the simulation to capture the formation of galaxies in the dense cluster environment and the surrounding filamentary structures. The inset panel shows soft band ($0.5-2.0\,\mathrm{keV}$) X-ray surface brightness within $2\,r_{500}$ of the potential minimum of the cluster. 

\section{Global properties}
\label{sec:screlations}
In this section we compare the global properties of the \textsc{c-eagle} sample at $z=0.1$ with low-redshift ($z\leq0.25$) observations. We also plot the groups and clusters from the periodic volumes of the \textsc{eagle} simulations. We label groups and clusters from the $100\,\mathrm{Mpc}$ volume run with the \textsc{eagle} reference model as `REF' and those from the $50\,\mathrm{Mpc}$ volume run with the AGNdT9 model as `AGNdT9'. To ensure a fair comparison with observational data we use quantities estimated via the mock X-ray analysis pipeline. However, we stress that estimated masses assume that the cluster is in hydrostatic equilibrium and that the X-ray estimated density and temperature profiles are good approximations of the true profiles. First, we test this assumption.

\subsection{Bias and scatter of estimated masses}
To examine the scatter and bias introduced by using estimated masses rather than the true masses, we plot the ratio of the estimated $M_{500}$ over $M_{500,\mathrm{true}}$ as a function of $M_{500,\mathrm{true}}$, where the `true' $M_{500}$ values are calculated via summation of particle masses that fall within the true $r_{500}$. To separate the effects of assuming hydrostatic equilibrium and estimating the profiles in an observational manner via the X-ray pipeline, we also fit the \citet{Vikhlinin2006} models to the true density and temperature profiles and label any quantity computed in this manner as `hse'.

\begin{figure*}
 \includegraphics[width=\textwidth]{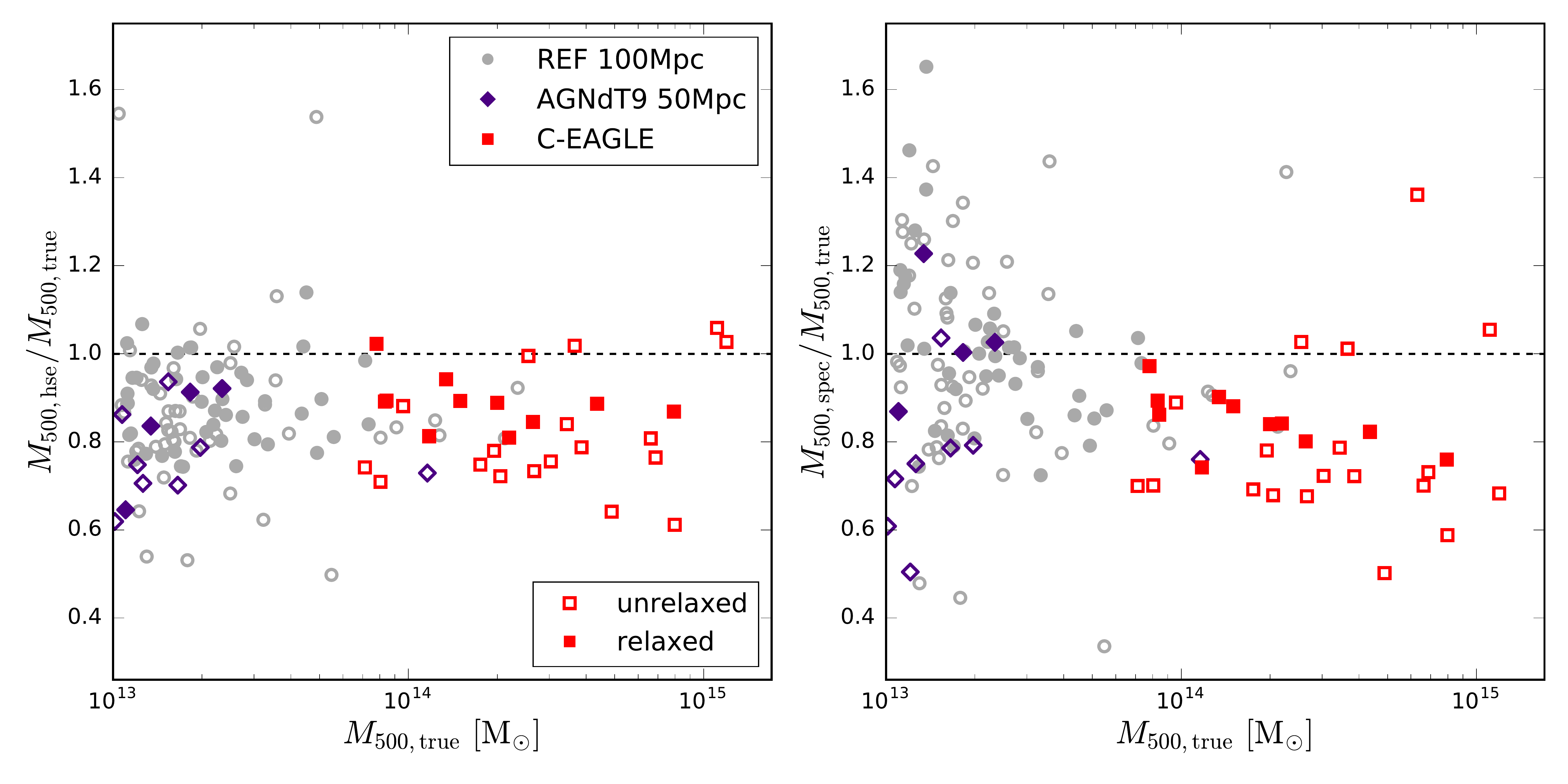}
 \caption{Ratio of estimated to true mass as a function of true mass at $z=0.1$ for the \textsc{c-eagle} clusters (red squares), as well as the groups and clusters from the \textsc{eagle} REF (grey circles) and AGNdT9 models (purple diamonds). The left panel shows hydrostatic mass estimates, calculated by fitting \citet{Vikhlinin2006} models to the true profiles, while the right panel shows X-ray spectroscopic mass estimates, calculated by fitting \citet{Vikhlinin2006} models to profiles estimated from the mock X-ray pipeline. Clusters that are defined as relaxed (unrelaxed) are shown by the filled (open) points. The dashed black line indicates no bias.}
 \label{fig:Mbias}
\end{figure*}

Fig. \ref{fig:Mbias} shows the scatter and bias introduced by estimating the mass for the \textsc{c-eagle} clusters as well as the REF and AGNdT9 groups and clusters. The assumption of hydrostatic equilibrium introduces a median bias $b_{\mathrm{hse}}=0.16\pm0.04$ for the \textsc{c-eagle} clusters, where $b_{\mathrm{hse}}=1-M_{500,\mathrm{hse}}/M_{500,\mathrm{true}}$ and the error is calculated by bootstrap resampling the data $10,000$ times. The assumption of hydrostatic equilibrium introduces a similar bias in the REF and AGNdT9 groups and clusters, which yield values of $b_{\mathrm{hse}}=0.14\pm0.02$ and $b_{\mathrm{hse}}=0.21\pm0.06$ respectively. We find that the bias introduced by assuming hydrostatic equilibrium is independent of mass. This is consistent with previous simulation work that has analyzed true profiles \citep{Nelson2014,Biffi2016,Henson2017}. For the \textsc{c-eagle} clusters, the profiles derived from the mock X-ray analysis produce a slightly larger bias with a median value $b_{\mathrm{spec}}=0.22\pm0.04$. Combined with the REF and AGNdT9 groups and clusters, it is clear that using the X-ray derived profiles increases the scatter in the estimated masses, with \textsc{c-eagle} yielding {\it r.m.s.} values of $\sigma_{\mathrm{hse}}=0.16$ and $\sigma_{\mathrm{spec}}=0.18$ for the hse and spec values respectively. Additionally, the spectroscopic bias appears to become mildly mass dependent for the mock X-ray pipeline mass estimates. This is consistent with previous simulation work that has made use of mock X-ray pipelines; \citet{LeBrun2014} saw an increase in scatter for low-mass ($<10^{13}\,\mathrm{M}_{\astrosun}$) groups and a median bias consistent with zero and \citet{Henson2017} found that the bias from mock X-ray pipelines showed a mild mass dependence.

Several of the \textsc{c-eagle} clusters have mass estimates that are more than $30\%$ discrepant from their true mass. To better understand the impact of the dynamical state of a cluster on the estimation of its mass, we classify each object as relaxed or unrelaxed. Theoretically there are many ways of defining whether a cluster is relaxed \citep[see][]{Neto2007,Duffy2008,Klypin2011,DuttonMaccio2014,Klypin2016}. In this work we define a cluster as being relaxed if 
\begin{equation}
 E_{\mathrm{kin},500,\mathrm{spec}}/E_{\mathrm{therm},500,\mathrm{spec}} < 0.1\,, \nonumber
\end{equation}
where $E_{\mathrm{kin},500,\mathrm{spec}}$ is the sum of the kinetic energy of the gas particles, with the bulk motion of the cluster removed, inside $r_{500}$ and $E_{\mathrm{therm},500,\mathrm{spec}}$ is the sum of the thermal energy of the gas particles within $r_{500}$. In all figures we denote relaxed (unrelaxed) clusters by solid (open) points. Using this criterion, $11$ of the $30$ \textsc{c-eagle} clusters are defined as relaxed. 

Selecting only relaxed (unrelaxed) \textsc{c-eagle} clusters we measure median mass biases $b_{\mathrm{hse}}=0.14\pm0.02\;(0.19\pm0.03)$ and $b_{\mathrm{spec}}=0.16\pm0.03\;(0.26\pm0.04)$. Thus, the mass estimates of relaxed clusters show a small decrease in the level of bias compared to the full sample, but the scatter in the mass estimate reduces significantly to $\sigma_{\mathrm{hse}}=0.06$ and $\sigma_{\mathrm{spec}}=0.06$. The bias of unrelaxed clusters increases slightly and the scatter about the median value increases to $\sigma_{\mathrm{hse}}=0.23$ and $\sigma_{\mathrm{spec}}=0.21$.
We see a similar trend for the REF and AGNdT9 groups and clusters. We also find that all \textsc{c-eagle} clusters with a mass estimate that is more than $30\%$ discrepant from its true mass have $E_{\mathrm{kin},500,\mathrm{spec}}/E_{\mathrm{therm},500,\mathrm{spec}} > 0.1$, demonstrating the impact of the dynamical state of the cluster on its estimated mass.

\subsection{Gas, stellar and black hole masses}
Theoretical work has shown that the gas and stellar content of clusters is largely controlled by stellar and particularly AGN feedback \citep{Voit2003,Bower2008,Fabjan2010,McCarthy2011,Planelles2013,LeBrun2014,Pike2014,Hahn2017,McCarthy2017}. Therefore, the gas, stellar and BH masses of the \textsc{c-eagle} clusters provide a test, orthogonal to usual tests of galaxy formation, of the feedback model and whether calibration on against galaxy properties alone leads to a reasonably realistic ICM.

\subsubsection{Stellar mass}
\begin{figure*}
 \includegraphics[width=\textwidth]{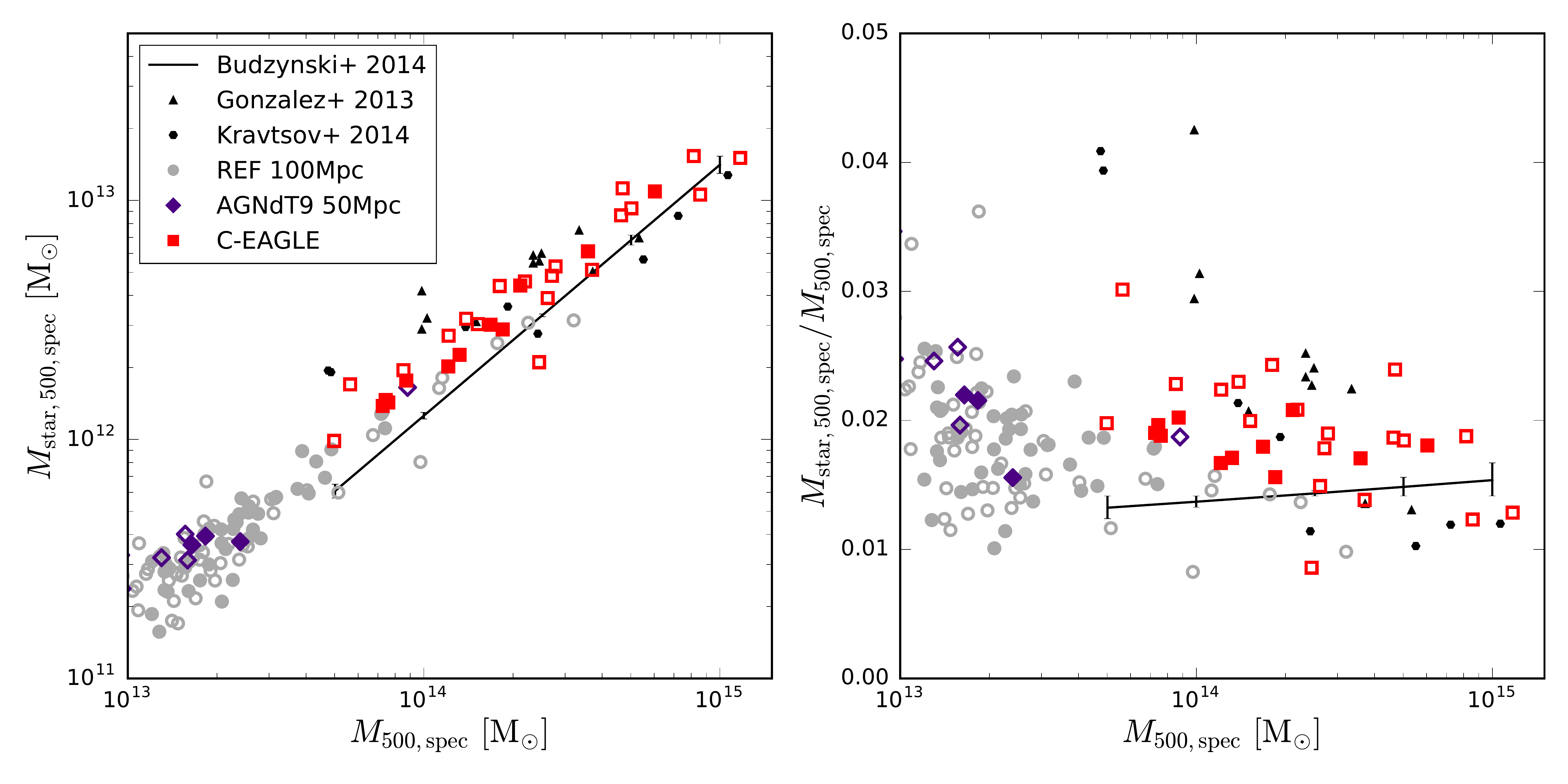}
 \caption{Integrated stellar mass (left panel) and stellar fraction (right panel) within $r_{500,\mathrm{spec}}$ as a function of estimated total mass at $z=0.1$ for the \textsc{c-eagle} clusters and the REF and AGNdT9 groups and clusters. Marker styles are the same as in Fig. \ref{fig:Mbias}. The black triangles and hexagons show the observational data from \citet{Gonzalez2013} and \citet{Kravtsov2014} respectively, and the black line with error bars shows the best-fit result from \citet{Budzynski2014} from stacking SDSS images.}
 \label{fig:MsM}
\end{figure*}

We plot the integrated stellar mass and stellar fraction within $r_{500,\mathrm{spec}}$ as a function of the estimated total mass at $z=0.1$, in the left and right panels of Fig. \ref{fig:MsM} respectively. To ensure a fair comparison, we only compare the \textsc{c-eagle}, REF and AGNdT9 samples against observations where the mass of the system has been estimated via high quality X-ray observations. The stellar mass-total mass relation within $r_{500,\mathrm{true}}$ is shown in Fig. 4 of Bah\'e et al. (2017). All of the observations and the simulations include the intracluster light in the stellar mass estimate and assume a \citet{Chabrier2003} initial mass function. First, we note that the observations do not appear to be consistent with each other. The results of \citet{Budzynski2014} have a lower normalization compared to the results of \citet{Kravtsov2014} and \citet{Gonzalez2013} at $M_{500}=10^{14}\,\mathrm{M}_{\astrosun}$, and the stellar fraction of \citet{Budzynski2014} increases slightly with total mass while \citet{Kravtsov2014} and \citet{Gonzalez2013} show a strongly decreasing stellar fraction with increasing total mass. This is most likely due to the different selection criteria and methods used in the observations. 

The \textsc{c-eagle} clusters provide a consistent extension into the high mass regime of the original periodic volumes. Estimating the cluster's mass from the X-ray hydrostatic analysis leads to an increased scatter about the relation compared to using the true mass, increasing from $\sigma_{\log_{10},\mathrm{true}}=0.07$ to $\sigma_{\log_{10},\mathrm{spec}}=0.10$.\footnote{We measure the scatter by fitting a power-law to the stellar mass-total mass relation for the full \textsc{c-eagle} sample, the relaxed subset and the unrelaxed subset. Further details of our method and a summary of the results for all scaling relations can be found in Appendix \ref{app:plf}.} If only relaxed systems are considered the scatter reduces substantially to $\sigma_{\log_{10},\mathrm{spec}}=0.03$. Taken together with groups and clusters from the periodic volumes, the \textsc{c-eagle} clusters show reasonable agreement with the observations. They reproduce the trend of the best-fit stellar mass-total mass relation of \citet{Budzynski2014}, but have a slightly higher normalization. In contrast, they reproduce the normalization of the \citet{Gonzalez2013} and \citet{Kravtsov2014} observations and are within the intrinsic scatter, but the trend with total mass is different. This is better seen in the stellar fraction-total mass relation, where the results of \citet{Gonzalez2013} and \citet{Kravtsov2014} show an increase in stellar fraction for lower mass objects, while the \textsc{c-eagle} clusters combined with the REF and AGNdT9 objects produce a roughly constant stellar mass fraction over two decades in mass. Overall, $\approx2\%$ of the total mass of the simulated groups and clusters consists of stars. This is consistent with previous numerical work, which has shown that the stars make up a few per cent of the cluster mass and the trend with mass is relatively mild \citep[e.g.][]{Planelles2013,Pike2014,Hahn2017}. The \textsc{eagle} model was calibrated to reproduce the observed galaxy stellar mass function of the field at low redshift, but dense cluster environments were not present in the calibration volumes due to their limited size. The level of agreement between the \textsc{c-eagle} clusters and the observations is therefore reassuring and demonstrates that the galaxy calibrations continue to work in the cluster regime. 

\subsubsection{BH masses}
\begin{figure*}
 \includegraphics[width=0.497\textwidth]{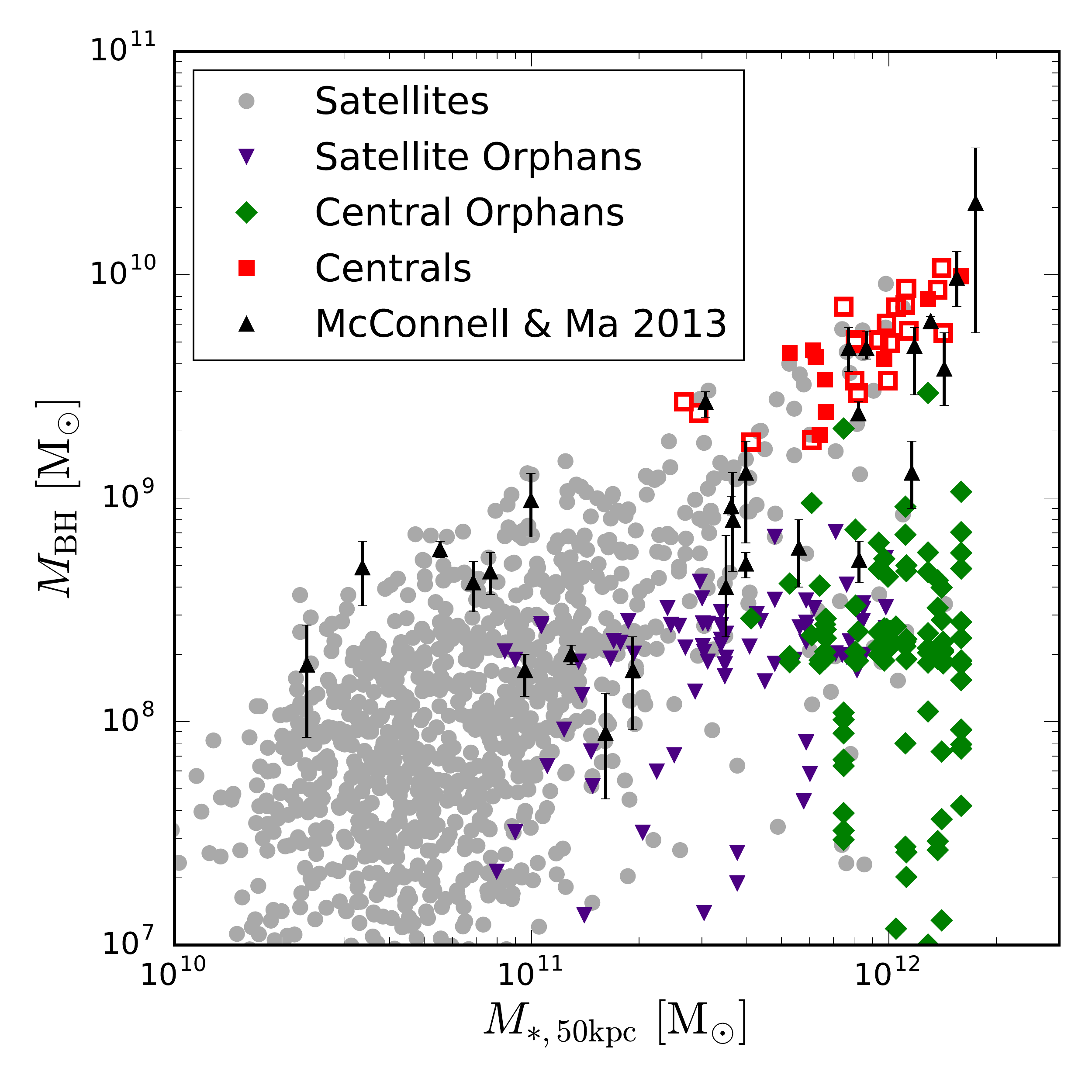}
 \includegraphics[width=0.497\textwidth]{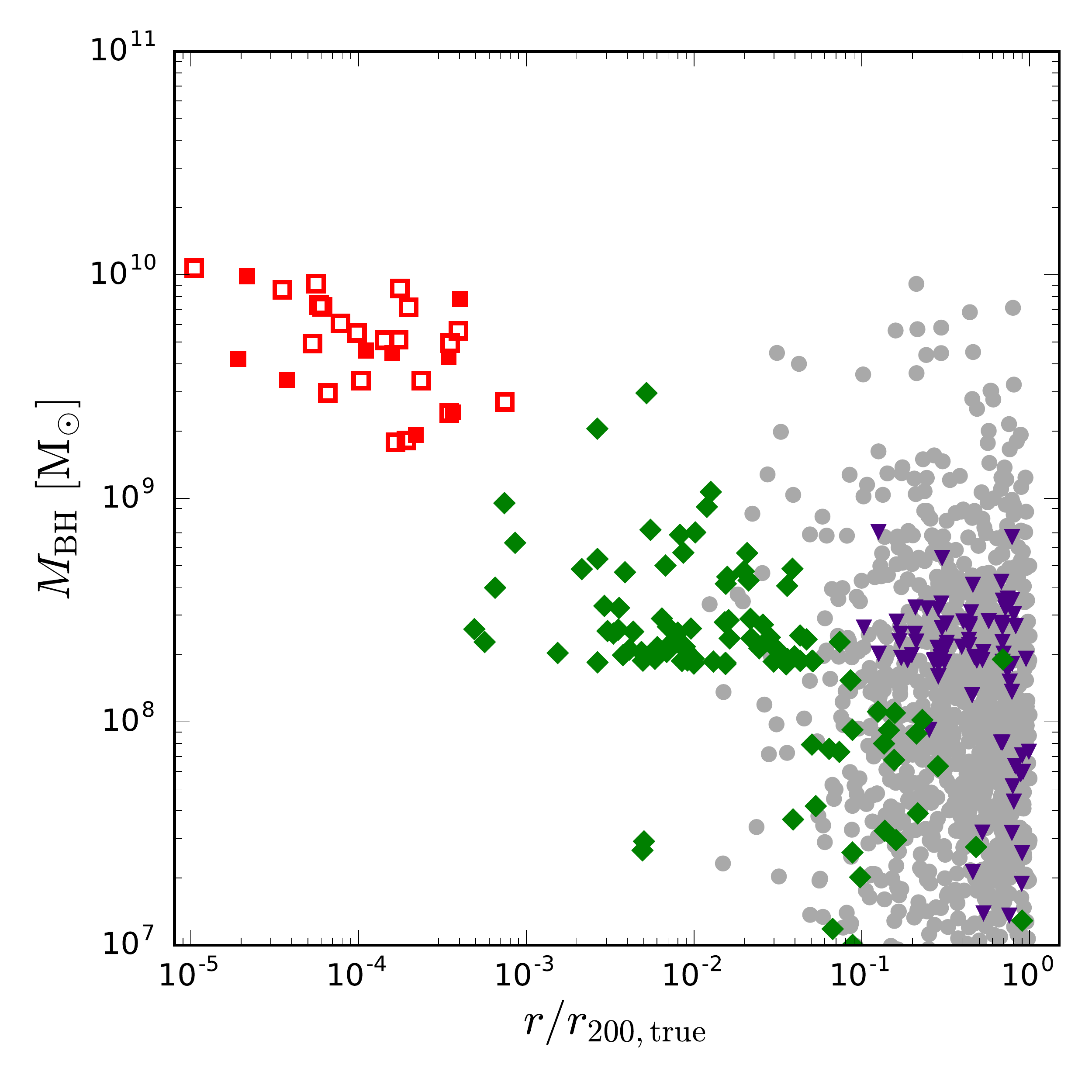}
 \caption{Black hole mass as a function of stellar mass within a $3$D $50\,\mathrm{kpc}$ aperture (left panel) and cluster-centric radius (right panel) for those BHs that fall within $r_{200}$ of a \textsc{c-eagle} cluster at $z=0.1$. The population is divided between centrals (red squares), satellites (grey circles), orphan black holes associated with the main halo (green diamonds) and orphans associated with satellites (purple triangles). For the centrals we denote whether the cluster is relaxed (unrelaxed) by filled (open) marker style. The black triangles show the observations of \citet{McConnellMa2013} for early type galaxies.}
 \label{fig:BHrel}
\end{figure*}

We examine the properties of the supermassive BHs that form in the \textsc{c-eagle} clusters in Fig. \ref{fig:BHrel}. In the left panel we plot BH mass as a function of stellar mass, calculated within a $3$D aperture of radius $50\,\mathrm{kpc}$, for those BHs with a mass $>10^{7}\,\mathrm{M}_{\astrosun}$ that fall within $r_{200,\mathrm{true}}$. In total, the \textsc{c-eagle} sample contains $1358$ BHs within $r_{200,\mathrm{true}}$. The aperture radius is a choice and we select $50\,\mathrm{kpc}$ as it mimics common observational choices. We decompose the bound population into centrals (those bound to the main halo) and satellites (those bound to subhaloes) as determined by the \textsc{subfind} algorithm \citep{Springel2001,Dolag2009}. We find that $90$ per cent of all BHs are defined as satellites. The central and satellite BHs form a continuous population on the BH mass-stellar mass relation. We compare the simulation relation to the observed early type galaxies from the compilation of \citet{McConnellMa2013}. We find good agreement with the observed relation, with the \textsc{c-eagle} clusters reproducing both the observed trend and normalization. As demonstrated in \citet{BoothSchaye2010}, the normalization of the BH mass-stellar mass relation is set by the feedback efficiency and the chosen value of $\epsilon_{\mathrm{f}}=0.15$ has been shown to work in low-resolution simulations of galaxies \citep{BoothSchaye2009}, clusters \citep{LeBrun2014} and in the \textsc{eagle} periodic volumes \citepalias{Schaye2015}.

Additionally, we define orphan BHs as those that are bound to either the main halo or a subhalo, but are not the most massive BH associated with it. We find that $\approx15$ per cent of BHs within $r_{200,\mathrm{true}}$ are classified as orphans. We futher classify the orphans as being associated either with the main halo or subhaloes, and find around $50$ per cent of the orphans are linked to the main halo. The number of orphans increases strongly with cluster mass and only subhaloes with a stellar mass $>10^{11}\,\mathrm{M}_{\astrosun}$ contain orphans. The \textsc{eagle} model allows BHs whose mass exceeds $100m_{\mathrm{gas}}$ to become dynamically independent and limits the distance that BHs less massive than this limit can drift towards the potential minimum at each time step. Therefore, when two haloes merge their BHs do not necessarily also instantaneously merge. In the right hand panel of Fig. \ref{fig:BHrel} we plot BH mass as a function of $r/r_{200,\mathrm{true}}$. The central BHs all lie within $0.1$ per cent of $r_{200,\mathrm{true}}$ from the potential minimum, even though they are all dynamically independent. No satellite BH is closer than $1$ per cent of $r_{200,\mathrm{true}}$. We find that $43$ orphans lie within $1\%$ of $r_{200,\mathrm{true}}$, which equates to a distance of $10-20\,\mathrm{kpc}$ from the potential minimum. Therefore, the \textsc{c-eagle} simulations predict that there is a population of BHs with a mass $>10^{8}\,\mathrm{M}_{\astrosun}$ inside the brightest cluster galaxies (BCG) that are not the central black hole. The orphans, both centrals and satellites, are likely the remnants of previous galaxy mergers, but we note that some may be identified with missed subhaloes as halo finders, such as \textsc{subfind}, have issues indentifying subhaloes in dense backgrounds \citep{Muldrew2011}. We leave a more detailed examination of the origins and properties of this orphan population to a future study.

\subsubsection{Gas mass}
\begin{figure*}
 \includegraphics[width=\textwidth]{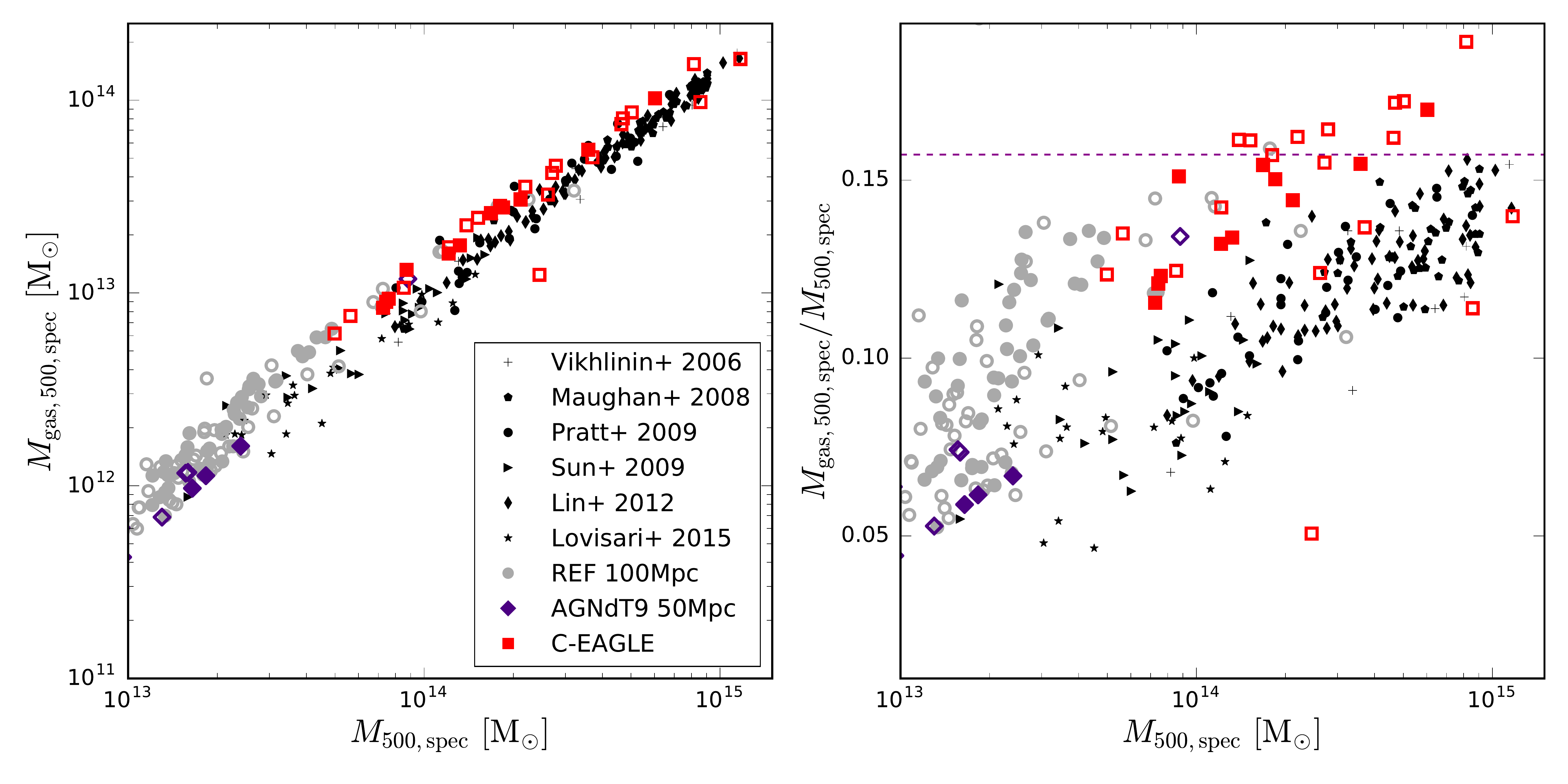}
 \caption{Gas mass as a function of estimated total mass within $r_{500,\mathrm{spec}}$ at $z=0.1$ for the \textsc{c-eagle} clusters and the REF and AGNdT9 groups and clusters. Marker styles are the same as in Fig. \ref{fig:Mbias}. The black pluses, pentagons, circles, right-facing triangles, thin diamonds and stars show the observational data from \citet{Vikhlinin2006}, \citet{Maughan2008}, \citet{Pratt2009}, \citet{Sun2009}, \citet{Lin2012} and \citet{Lovisari2015} respectively. The left panel shows the gas mass, whereas the right panel shows gas fractions with the magenta dashed line showing the universal baryon fraction, $\Omega_{\mathrm{b}}/\Omega_{\mathrm{M}}=0.157$.}
 \label{fig:MgM}
\end{figure*}

In Fig. \ref{fig:MgM} we plot the gas mass and the gas fraction within $r_{500,\mathrm{spec}}$ as a function of the estimated total mass at $z=0.1$, in the left and right panels respectively. Again, we compare against the groups and clusters from the REF and AGNdT9 volumes and observational data. The \textsc{c-eagle} clusters provide a consistent extension to the objects from the periodic volumes. The \textsc{c-eagle} clusters reproduce the observed trend with halo mass, but they are too gas rich and lie along the top of the observed scatter of the gas mass-total mass relation. Although we selected the AGNdT9 \textsc{eagle} model as it produced a better match to the observed gas fractions of low-mass ($<10^{13.5}\,\mathrm{M}_{\astrosun}$) groups, the subgrid model was not calibrated at cluster scales. Therefore, the gas mass and fractions are a prediction of a model calibrated for field galaxy formation. Many previous numerical simulations yield a better match to the observed gas fraction relation \citep{Planelles2013,LeBrun2014,Pike2014,McCarthy2017}. However, these simulations have significantly lower resolution and use traditional SPH, which has been shown by \citet{Schaller2015} to have an impact on group scales. We note that one of the \textsc{c-eagle} clusters is undergoing a major merger at $z=0.1$, displacing the gas from the potential minimum and leading to an estimated gas mass well below the observed relation.

We also plot the gas fractions as the dynamic range of the gas mass-total mass relation hides some of the discrepancies. As shown in \citetalias{Schaye2015}, the low-mass $(<10^{13.5}\,\mathrm{M}_{\astrosun})$ groups of the periodic volumes show a significant difference in the gas fraction for the two AGN calibrations, with the increased AGN heating temperature of the AGNdT9 model producing gas fractions that are systematically lower by $\approx30$ per cent. However, at cluster scales $(>10^{14}\,\mathrm{M}_{\astrosun})$ the runs with different heating temperatures converge towards to the same relation, although we currently suffer from a small sample size. Gas fractions for some of the \textsc{c-eagle} clusters appear to be greater than the universal baryon fraction, $f_{\mathrm{b}}\equiv\Omega_{\mathrm{b}}/\Omega_{\mathrm{M}}=0.157$. This is due to using an X-ray estimated mass that is biased low compared to the true mass.

\begin{figure}
 \includegraphics[width=\columnwidth,height=8cm]{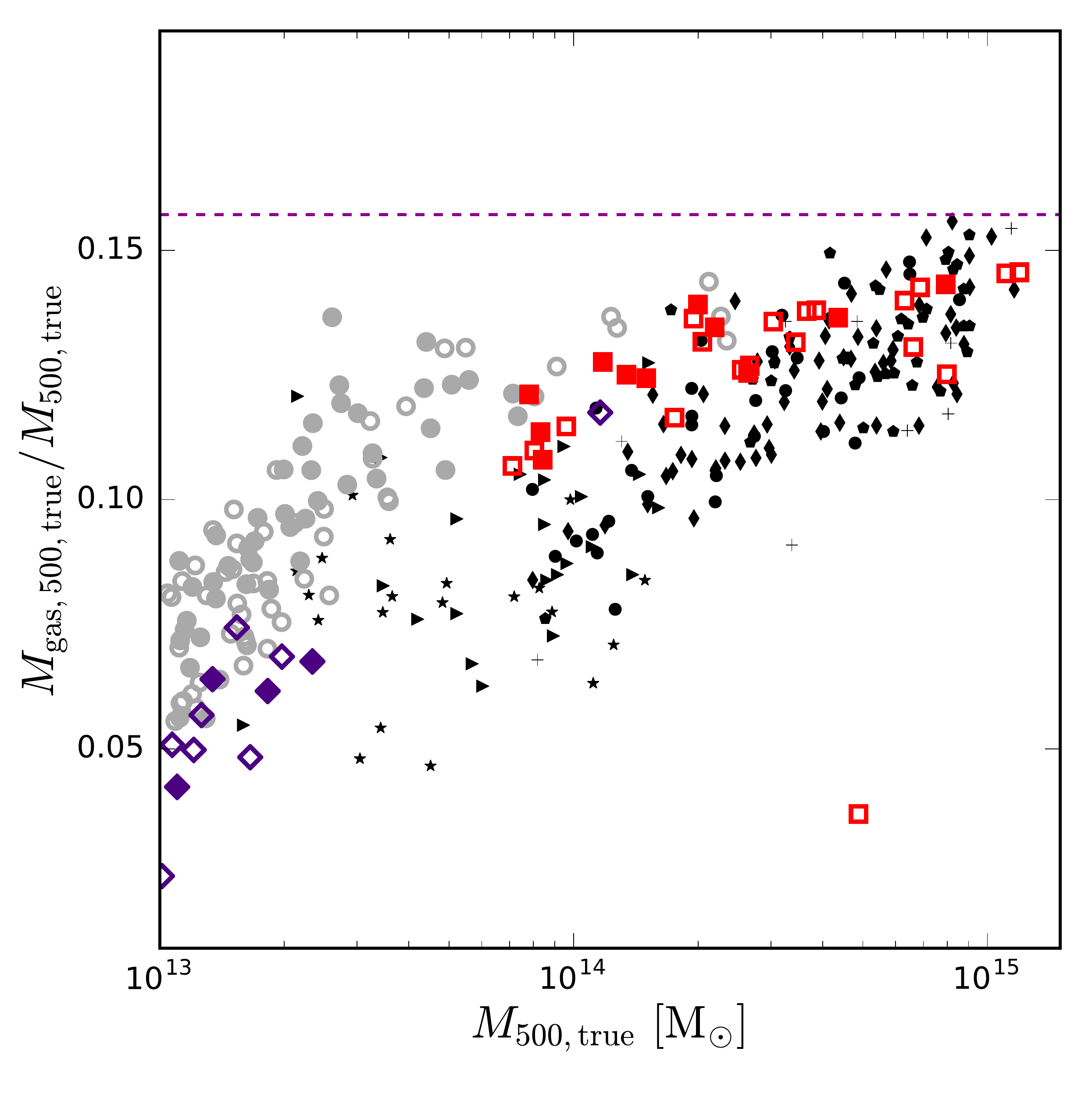}
 \caption{True gas fraction as a function of true total mass, both within $r_{500,\mathrm{spec}}$, at $z=0.1$ for the \textsc{c-eagle} clusters and the REF and AGNdT9 groups and clusters. Marker styles and observations are the same as in Fig. \ref{fig:MgM}.}
 \label{fig:MgM_true}
\end{figure}

To remove the impact of using an estimated mass, we plot the true gas fraction as a function of $M_{500,\mathrm{true}}$ in Fig. \ref{fig:MgM_true}. The use of true mass significantly reduces the scatter in the relation and all of the simulated clusters are now below the universal fraction, with the most massive clusters reaching $\approx90$ per cent of the universal fraction. However, the simulated clusters and groups above a mass of $M_{500,\mathrm{spec}}\geq10^{13.5}\,\mathrm{M}_{\astrosun}$ are still too gas rich. Previous work has shown that the majority of gas expulsion occurs in the progenitors of a halo at high redshift \citep{McCarthy2011}. These progenitors form at an earlier epoch for more massive objects, when the Universe was denser, making their potentials deeper. The higher-than-observed gas fractions of the groups and clusters suggest that the AGN feedback in the \textsc{eagle} model is not efficient enough at removing gas from the deeper potentials of the progenitors of massive groups and clusters ($>10^{14}\,\mathrm{M}_{\astrosun}$). This also leads to some overcooling and larger than observed BCGs (see right panel of Fig. 4 of \citealt{Bahe2017}). We note that the true gas fractions for rich clusters ($10^{15}\,\mathrm{M}_{\astrosun}$) appear to be in agreement with the observational data, however, this is clearly no longer a fair comparison.

\subsection{X-ray and Sunyaev-Zel'dovich properties}
Having shown that the \textsc{c-eagle} clusters have stellar and BH masses in reasonable agreement with the observed relations but are too gas rich, we now examine the X-ray and SZ observable properties at $z=0.1$ produced by the mock X-ray pipeline.

\subsubsection{Spectroscopic temperature-total mass}
\begin{figure*}
 \includegraphics[width=\textwidth]{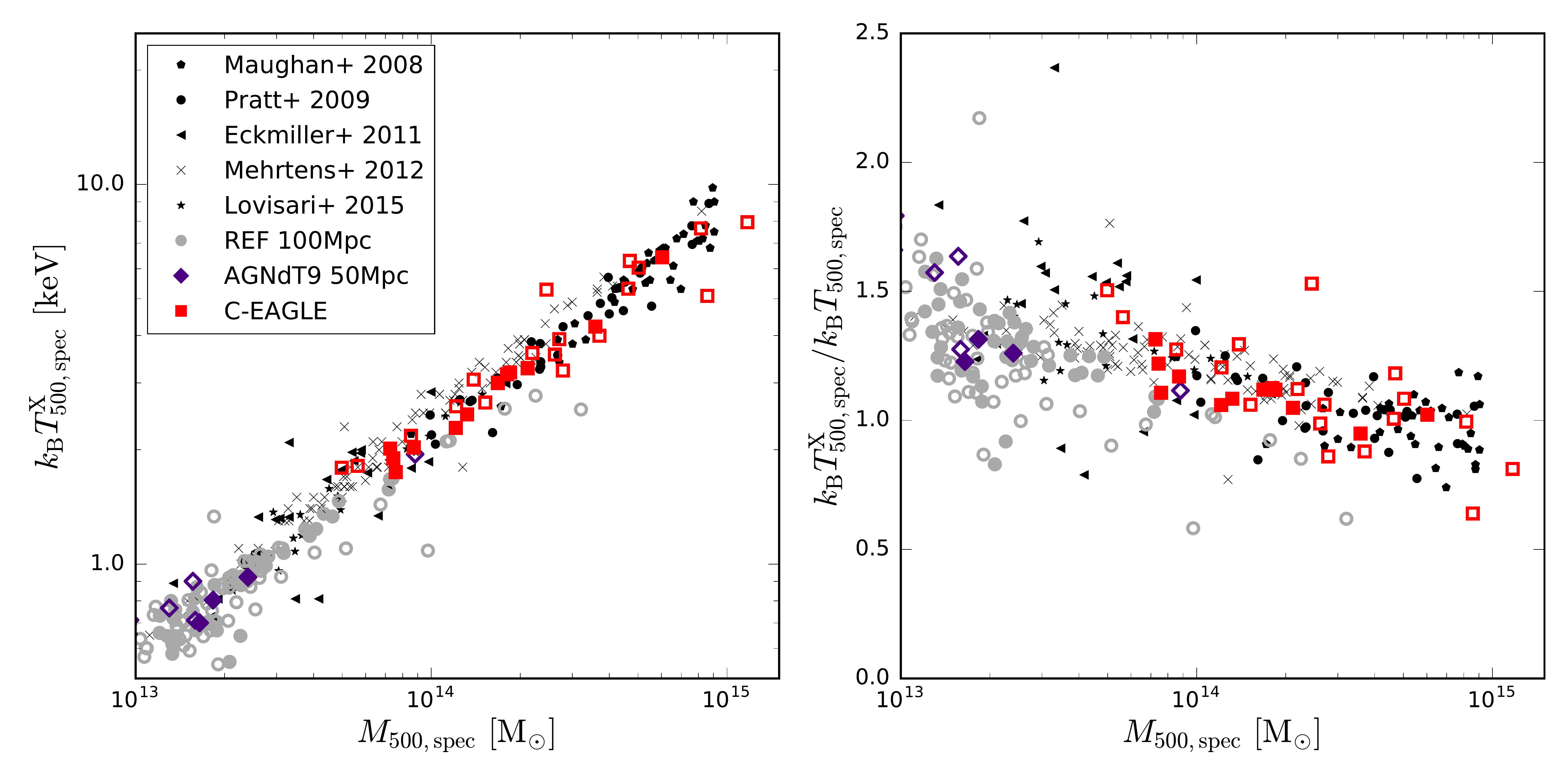}
 \caption{Spectroscopic temperature measured within $r_{500,\mathrm{spec}}$ as a function of estimated total mass at $z=0.1$ for the \textsc{c-eagle} clusters and the REF and AGNdT9 groups and clusters. Marker styles are the same as in Fig. \ref{fig:Mbias}. The black pentagons, circles, left-facing triangles, crosses and stars show observational data from \citet{Maughan2008}, \citet{Pratt2009}, \citet{Eckmiller2011}, \citet{Mehrtens2012} and \citet{Lovisari2015} respectively. The left panel shows the temperature and the right panel shows the temperature normalized by the virial temperature to approximately remove the mass dependence.}
 \label{fig:TxM}
\end{figure*}

In the left panel of Fig. \ref{fig:TxM} we plot the spectroscopic X-ray temperature, $k_{\mathrm{B}}T^{\mathrm{X}}_{500,\mathrm{spec}}$, measured within $r_{500,\mathrm{spec}}$ as a function of the estimated total mass. The \textsc{c-eagle} clusters and the groups and clusters from the REF and AGNdT9 volumes are a good match to the observational data, yielding a linear power law relation over two decades in mass. The \textsc{c-eagle} simulations have a scatter of $\sigma_{\log_{10}}=0.06$, which is similar to the value of $\sigma=0.05$ found by \citet{Eckmiller2011} for the \textsc{hiflugcs} sample. Though we calculate properties in a manner consistent with observational techniques, we do not make any attempt to account for observational selection effects. The clusters with the largest scatter are all unrelaxed and these objects may not be selected in X-ray samples due to the $n_{\mathrm{H}}^{2}$ dependence of the emission, which results in a bias towards more relaxed systems.

In the right panel of Fig. \ref{fig:TxM} we have removed the expected mass dependence of the relation that results from hydrostatic equilibrium by dividing each system by its virial temperature, which we calculate as
\begin{equation}
 k_{\mathrm{B}}T_{500,\mathrm{spec}}\equiv GM_{500,\mathrm{spec}}\mu m_{\mathrm{p}}/2r_{500,\mathrm{spec}}\:,
\end{equation}
where $k_{\mathrm{B}}$ is the Boltzmann constant, $G$ is Newton's gravitational constant, $\mu=0.59$ is the mean molecular weight of the gas and $m_{\mathrm{p}}$ is the proton mass. This allows us to examine the impact of non-gravitational processes, such as cooling and feedback. 

The most massive \textsc{c-eagle} clusters have temperature ratios that are close to one, showing that their temperatures are close to the expected virial values. As the mass of the object decreases the temperature increases relative to the virial temperature, being $50$ per cent higher than expected for the $10^{13}\,\mathrm{M}_{\astrosun}$ groups in the REF and AGNdT9 samples. This demonstrates that the feedback is able to heat and expel gas more efficiently in lower mass objects. This is consistent with previous simulation work \citep{Short2010,McCarthy2010,Planelles2014,LeBrun2014,Hahn2017}. However, in \citet{LeBrun2014} an increase in the AGN heating temperature produced a flatter normalized temperature relation (see the right hand panel in their Fig. 2), a difference that we do not see between the REF and AGNdT9 models. This may be due to resolution as the \textsc{c-eagle} clusters achieve a factor $~500\times$ better mass resolution, which reduces the energy per AGN feedback event and may reduce the impact of feedback on the cluster properties.

\subsubsection{X-ray luminosity-total mass}
\begin{figure}
 \includegraphics[width=\columnwidth]{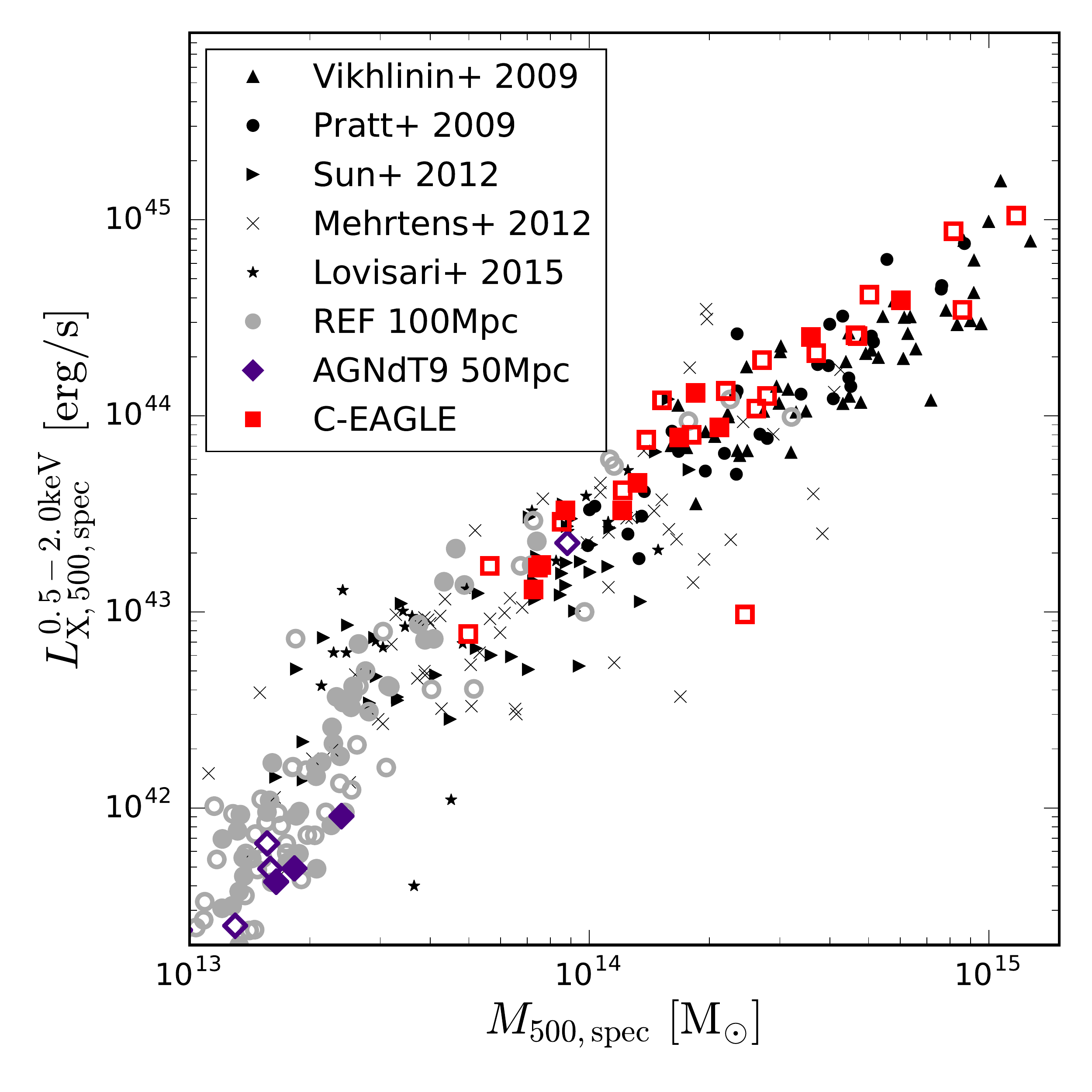}
 \caption{Soft band X-ray luminosity within $r_{500,\mathrm{spec}}$ as a function of estimated total mass at $z=0.1$ for the \textsc{c-eagle} clusters and the REF and AGNdT9 groups and clusters. Marker styles are the same as in Fig. \ref{fig:Mbias}. The black triangles, circles, right-facing triangle, crosses and stars show observational data from \citet{Vikhlinin2009}, \citet{Pratt2009}, \citet{Sun2012}, \citet{Mehrtens2012} and \citet{Lovisari2015} respectively.}
 \label{fig:LxM}
\end{figure}

We plot the soft band $(0.5-2.0\,\mathrm{keV})$ X-ray luminosity, $L^{0.5-2.0\,\mathrm{keV}}_{\mathrm{X},500,\mathrm{spec}}$, within $r_{500,\mathrm{spec}}$ as a function of the estimated total mass at $z=0.1$ in Fig. \ref{fig:LxM}. The \textsc{c-eagle} clusters show reasonable agreement with the observational data, reproducing the observed trend. The normalization of the simulated trend is marginally high compared to the observed relation, but still within the scatter, and this is due to the clusters being too gas rich. We note that the merging cluster has a significantly decreased X-ray luminosity for its mass. The complete \textsc{c-eagle} sample has a scatter of $\sigma_{\log_{10}}=0.30$, which is larger than the values of $\sigma_{\log_{10}}=0.17$ and $\sigma_{\log_{10}}=0.25$ found for the \textsc{REXCESS} \citep{Pratt2009} and \textsc{HIFLUGCS} \citep{Lovisari2015} samples respectively. However, we stress that we do not attempt to account for selection effects. If we select only relaxed clusters the scatter reduces to $\sigma_{\log_{10}}=0.11$. The \textsc{c-eagle} clusters are consistent with the clusters of the REF and AGNdT9 samples, with the models producing negligible difference for the X-ray luminosity-mass relation.

At group scales we find a mild break in the power-law relation. The \textsc{eagle} model reproduces the observed luminosity-mass relation over two decades in mass and 4 decades in luminosity. Previous numerical works that include AGN feedback and reproduce the observed gas fraction relation also reproduce the observed X-ray luminosity-total mass relation \citep[e.g.][]{LeBrun2014,Pike2014,Planelles2014,McCarthy2017}, while those without, or ineffective, AGN feedback \citep{Biffi2014,Hahn2017}  tend to over-estimate the normalization of the relation. We note that the scatter increases for observed low-mass groups, but the scatter in the simulated systems does not. However, this increase in the scatter of observed groups may simply reflect the increasing challenge of measuring the total X-ray emission from low-mass groups, rather than a failure of the simulations to reproduce the scatter of the observed relation.

\subsubsection{Metallicity-spectroscopic temperature}
\begin{figure}
 \includegraphics[width=\columnwidth]{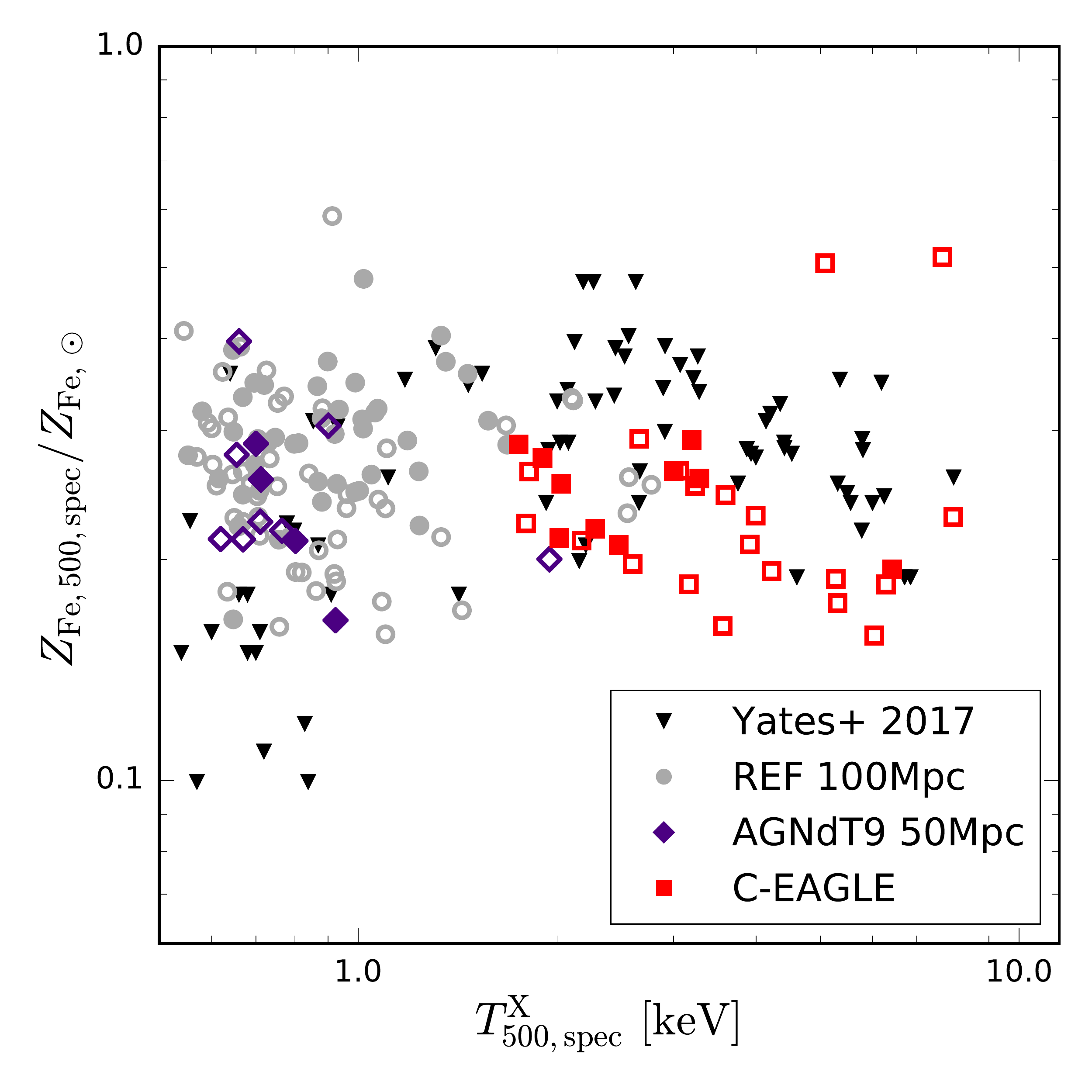}
 \caption{Mass-weighted iron abundance measured within $r_{500,\mathrm{spec}}$ as a function of spectroscopic temperature at $z=0.1$ for the \textsc{c-eagle} clusters and the REF and AGNdT9 groups and clusters. Marker styles are the same as in Fig. \ref{fig:Mbias}. The black triangles show the homogenized observational data from \citet{Yates2017}.}
 \label{fig:ZxTx}
\end{figure}

In Fig. \ref{fig:ZxTx} we plot the `mass-weighted' iron abundance within $r_{500,\mathrm{spec}}$ as a function of the spectroscopic temperature at $z=0.1$. Following \citet{Yates2017}, we estimate the mass-weighted iron abundance via
\begin{equation}
 Z_{\mathrm{Fe},500,\mathrm{spec}} = \frac{\int_{0}^{r_{500,\mathrm{spec}}}Z_{\mathrm{Fe,spec}}\rho_{\mathrm{gas,spec}}r^{2}dr}{\int_{0}^{r_{500,\mathrm{spec}}}\rho_{\mathrm{gas,spec}}r^{2}dr}\:,
\end{equation}
where $Z_{\mathrm{Fe,spec}}$ and $\rho_{\mathrm{gas,spec}}$ are the iron abundance and gas density profiles, respectively, estimated from single temperature fits to the summed particle spectra for $25$ $3$D cluster-centric radial bins and we integrate over those bins that fall within $r_{500,\mathrm{spec}}$. We compare the simulated samples against consolidated data taken from \citet{Yates2017}; see the references therein for more details on the homogenization process and the different samples. If more than one estimate for the observed mass-weighted metallicity or temperature was present for a system the mean value was taken. We have scaled both the simulated and observational data to the solar abundances of \citet{Asplund2009}.

The \textsc{c-eagle} clusters show reasonable agreement with the observed metallicity-temperature relation, with a median metallicity of $Z^{\mathrm{med}}_{\mathrm{Fe},500,\mathrm{spec}}=0.23\,\mathrm{Z}_{\astrosun}$ and an {\it r.m.s.} scatter of $\sigma=0.09$. The REF and AGNdT9 groups clusters are also in agreement with the data, with median metallicities of $Z^{\mathrm{med}}_{\mathrm{Fe},500,\mathrm{spec}}=0.26\,\mathrm{Z}_{\astrosun}$ and $Z^{\mathrm{med}}_{\mathrm{Fe},500,\mathrm{spec}}=0.22\,\mathrm{Z}_{\astrosun}$ and scatters of $\sigma=0.07$ and $\sigma=0.06$ respectively. Overall, the \textsc{eagle} model produces a relatively flat mass-weighted metallicity-spectroscopic temperature relation from low-mass groups to rich clusters. Reproducing the global metallicities of groups and clusters has been a challenge for previous simulation works, with \citet{Martizzi2016} under-predicting the total metal content of clusters, \citet{Planelles2014} over-predicting the metallicity of more massive clusters and \citet{Yates2017} over-predicting the metallicity of groups. Recent observational \citep{McDonald2016,Mantz2017} and numerical \citep{Biffi2017} results demonstrate that the global cluster metallicity shows very little evolution, suggesting that simulations must carefully model the early galaxy formation processes to reproduce the observed cluster metallicity.

\subsubsection{Y-total mass relations}
\begin{figure*}
 \includegraphics[width=0.497\textwidth]{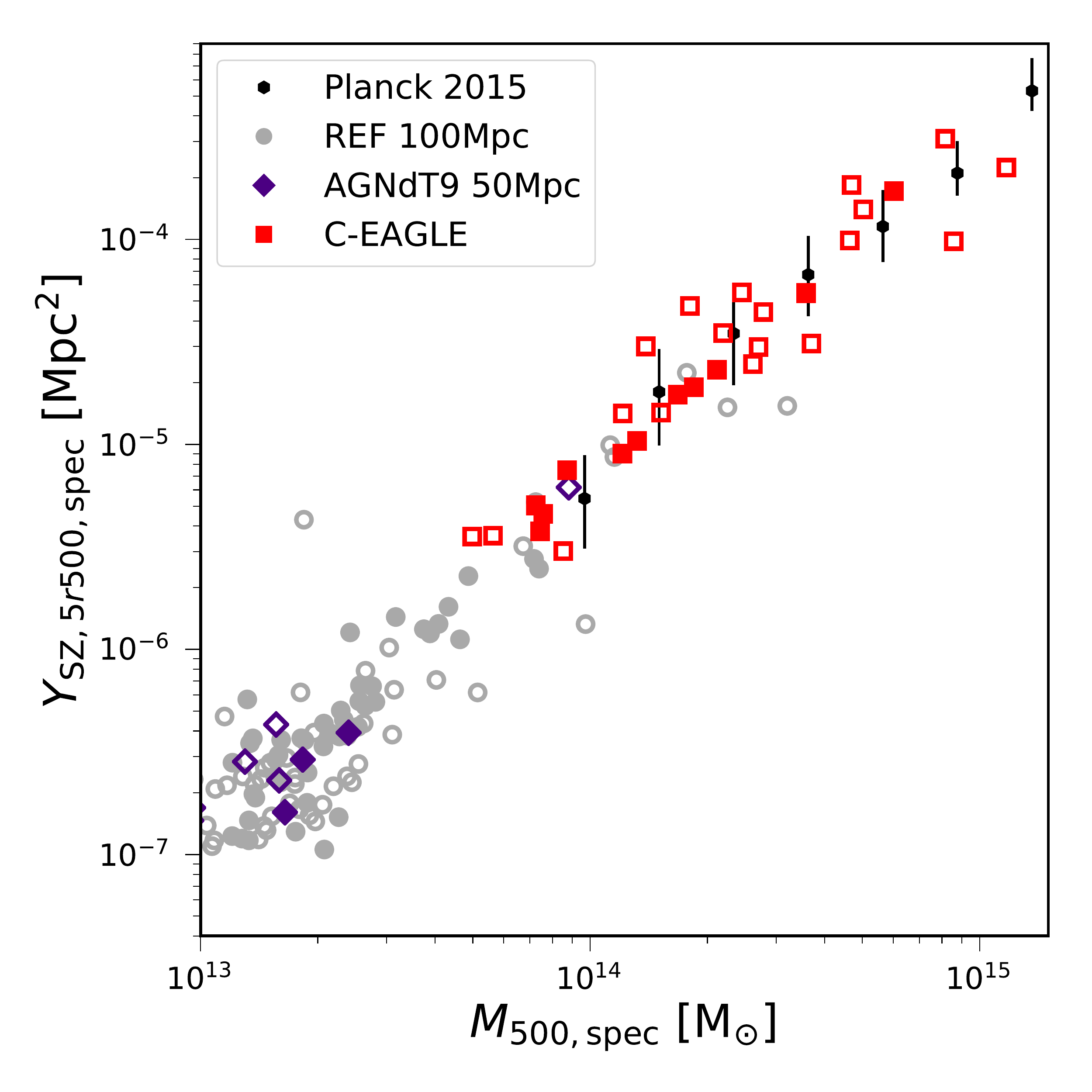}
 \includegraphics[width=0.497\textwidth]{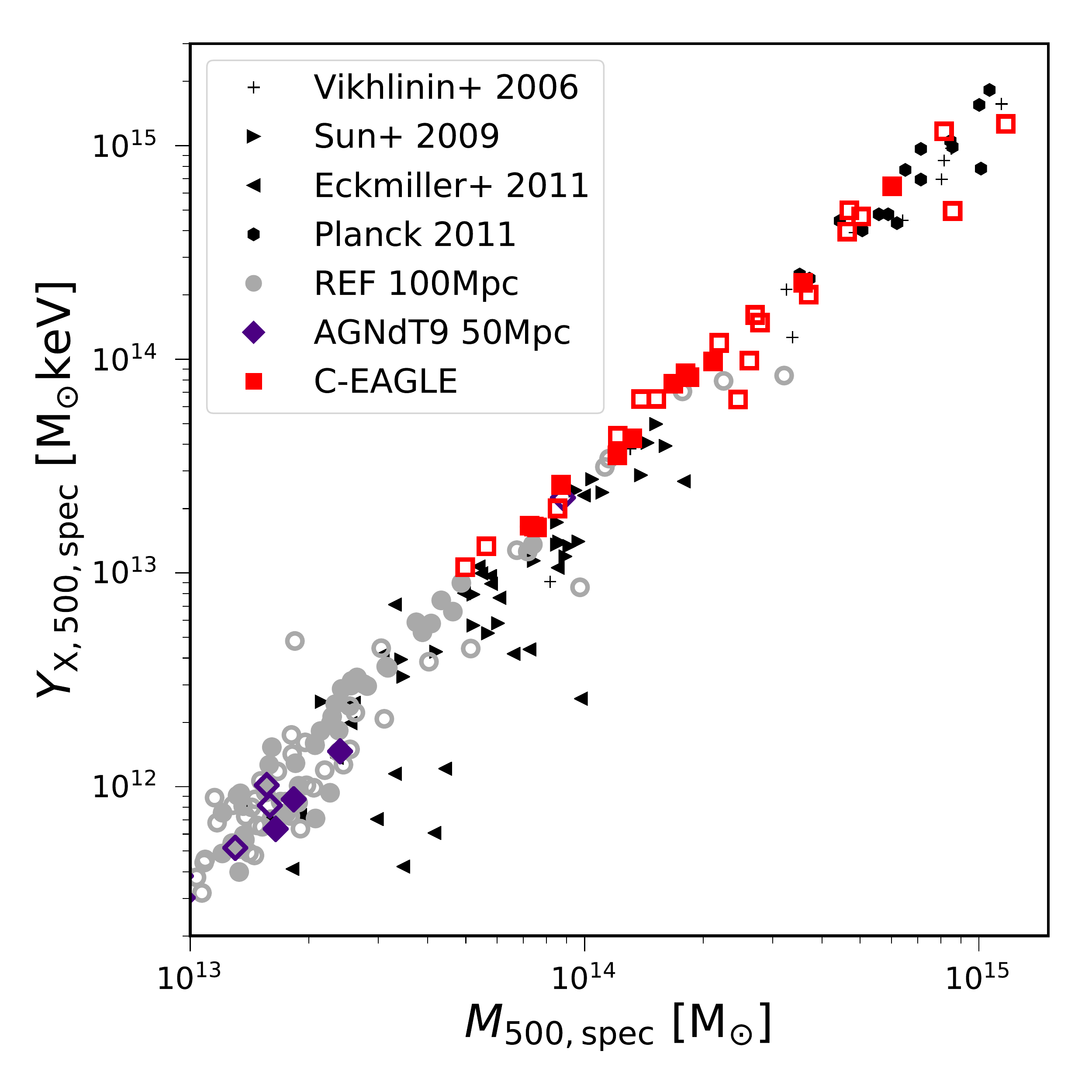}
 \caption{Sunyaev-Zel'dovich signal within $5r_{500,\mathrm{spec}}$ (left panel) and $Y_{\mathrm{X}}$ within $r_{500,\mathrm{spec}}$ (right panel) as a function of estimated total mass within $r_{500,\mathrm{spec}}$ at $z=0.1$ for the \textsc{c-eagle} clusters and the REF and AGNdT9 groups and clusters. Marker styles are the same as in Fig. \ref{fig:Mbias}. In the left panel the black hexagons show the median relation from the second \textit{Planck} SZ catalogue \citep{PlanckSZ2015}, with the error bars denoting $68\%$ of the observed clusters. In the right hand panel the black pluses, right-facing triangles, left-facing triangles and hexagons show observational data from \citet{Vikhlinin2006}, \citet{Sun2009}, \citet{Eckmiller2011} and \citet{Planck2011} respectively.}
 \label{fig:YM}
\end{figure*}

The SZ signal, $Y_{\mathrm{SZ}}$, is a measure of the total thermal energy content of the ICM, and is thought to be relatively insensitive to the details of the baryonic physics within the cluster volume \citep{daSilva2004,Nagai2006}, although see \citet{LeBrun2014}. The X-ray analogue of the SZ signal, $Y_{\mathrm{X}}$, was first proposed by \citet{Kravtsov2006} and we define it as the product of the core-excised spectroscopic temperature and the gas mass. The difference between the two quantities is that $Y_{\mathrm{SZ}}$ is dependent on the mass-weighted temperature, whereas $Y_{\mathrm{X}}$ is dependent on the core-excised spectroscopic temperature. We plot $Y_{\mathrm{SZ}}$ within $5r_{500,\mathrm{spec}}$ and $Y_{\mathrm{X}}$ within $r_{500,\mathrm{spec}}$ as a function of the estimated total mass within $r_{500,\mathrm{spec}}$ at $z=0.1$ in Fig. \ref{fig:YM}. For the SZ signal we compare against clusters from the second \textit{Planck} SZ catalogue \citep{PlanckSZ2015} that are not infrared contaminated, have a neural quality $>0.4$ (the recommended quality threshold), have a mass estimate and are at a redshift of $z<0.25$. This yields an observed sample of over $600$ clusters and we remove the redshift dependence of the fluxes by scaling by the square of the angular diameter distance and applying a self-similar scaling of $E^{-2/3}(z)$, where $E(z)\equiv H(z)/H_{0}=\sqrt{\Omega_{\mathrm{M}}(1+z)^{3}+\Omega_{\mathrm{\Lambda}}}$. We then binned the clusters into $\log_{10}$ mass bins of width $0.2$ dex and calculated the median value and $1\sigma$ percentiles.

The \textsc{c-eagle} clusters provide a consistent extension to the groups and clusters from the periodic volumes. The different AGN calibrations produce negligible difference, even at low-mass group scales. The simulations show good agreement with the observed trend producing a tight power-law relation from low-mass groups to rich clusters, with the greatest outliers being unrelaxed clusters. The normalization of the $Y_{\mathrm{X}}$ relation is marginally higher than the observed relation. This is due to the clusters being too gas rich. The $Y_{\mathrm{SZ}}$ relation does not suffer from the problem of clusters being too gas rich as we measure it within a larger aperture, and the gas fraction averaged within a sphere with such a large radius will tend towards the universal baryon fraction. Our results are consistent with previous numerical work, which have reproduced the observed $Y$-total mass scaling relations independent of hydrodynamical method and subgrid model \citep[e.g.][]{Kay2012,Battaglia2012,LeBrun2014,Pike2014,Planelles2014,Yu2015,Gupta2016,Planelles2017}. For the $Y_{\mathrm{X}}$ relation the \textsc{c-eagle} sample has a scatter of $\sigma_{\log_{10}}=0.10$, which is similar to the intrinsic scatter of $\sigma_{\log_{10}}=0.12$ in the \textsc{hiflugcs} sample \citep{Eckmiller2011}. If we select relaxed clusters we observe a tighter power-law with a scatter of $\sigma_{\log_{10}}=0.04$, which is consistent to the scatter of $\sigma_{\log_{10}}=0.04$ measured by \citet{Arnaud2007} for a sample of X-ray selected clusters. The $Y_{\mathrm{SZ}}$-total mass relation shows similar behaviour, with a scatter of $\sigma_{\log_{10}}=0.18$ for the full sample reducing to $\sigma_{\log_{10}}=0.06$ for the relaxed sample.

\subsection{Summary}
We have compared the global properties of the \textsc{c-eagle} clusters against observational data and the groups and clusters from two of the periodic volumes from the original \textsc{eagle} project. We found that the \textsc{c-eagle} clusters provided a consistent high-mass extension to the periodic volumes for all global scaling relations. We examined the bias introduced by using estimated masses rather than true masses and demonstrated that the assumption of hydrostatic equilibrium introduces a mass independent bias of $b_{\mathrm{hse}}=0.16$. The use of X-ray estimated profiles increases the bias slightly, leads to greater scatter in the mass estimate and results in a mild mass dependence, with the bias increasing for more massive clusters. Overall the \textsc{c-eagle} clusters show reasonable agreement with the observed global scaling relations, reproducing the stellar mass, spectroscopic temperature, X-ray luminosity and SZ signal as a function of total mass relations, the BH mass-stellar mass relation and the global metallicity-spectroscopic temperature relation. The exception is the gas mass-total mass relation where the observed trend is reproduced, but the normalization is too high and the \textsc{c-eagle} clusters lie at the top of the intrinsic scatter of the observational data. This impacts the $Y_{\mathrm{X}}$-total mass relation and its normalization is slightly too high, but the observed trend is reproduced. The higher than observed gas fractions suggest that the AGN feedback is not efficient enough at ejecting gas from the deeper potentials of the progenitors of massive haloes that form at an earlier epoch. Defining a relaxed sample based on the ratio of the kinetic and thermal energy of the hot gas in the ICM, we found that for many relations the scatter is significantly reduced when only considering relaxed clusters.

\section{Hot gas profiles}
\label{sec:gasprofs}
We now examine the hot gas radial profiles of the \textsc{c-eagle} clusters to better characterize its distribution and the impact of the AGN feedback model. To make quantitative comparisons to observational data we must compare like-with-like. The \textsc{c-eagle} sample has a median mass of $M_{500,\mathrm{spec}}=1.80\times10^{14}\,\mathrm{M}_{\astrosun}$ and median radius of $r_{500,\mathrm{spec}}=0.88\,\mathrm{Mpc}$. We compare the profiles from the  mock X-ray pipeline to the REXCESS cluster sample \citep{Bohringer2007}, but we cut the observational sample such that all clusters have $M_{500}>10^{14}\,\mathrm{M}_{\astrosun}$, $z<0.25$ and the median mass of the sample is $M_{500}=2.1\times10^{14}\,\mathrm{M}_{\astrosun}$. Matching the median masses of the samples ensures a fairer comparison. However, we do not make any attempt to correct for selection effects. In this section we calculate dimensionless profiles by dividing by the appropriate quantity, i.e. $\rho_{\rm{crit}}$, $k_{\rm{B}}T_{500,\mathrm{spec}}$, $P_{500,\mathrm{spec}}$ or $K_{500,\mathrm{spec}}$, where we define these quantities as
\begin{equation}
 \rho_{\rm{crit}}(z)\equiv E^2(z)\frac{3H^2_0}{8\pi G}\,,
\end{equation}
\begin{equation}
 k_{\rm{B}}T_{500,\mathrm{spec}}=\frac{GM_{500,\mathrm{spec}}\mu m_{\rm{p}}}{2r_{500,\mathrm{spec}}}\,,
\end{equation}
\begin{equation}
 P_{500,\mathrm{spec}}=500f_{\rm{b}}k_{\rm{B}}T_{500,\mathrm{spec}}\frac{\rho_{\rm{crit}}}{\mu m_{\rm{p}}}\,,
\end{equation}
\begin{equation}
 K_{500,\mathrm{spec}}=\frac{k_{\rm{B}}T_{500,\mathrm{spec}}}{\left(500f_{\rm{b}}(\rho_{\rm{crit}}/\mu_{\rm{e}}m_{\rm{p}})\right)^{2/3}}\,,
\end{equation}
where $H_0$ is the Hubble constant and $\mu_{\rm{e}}=1.14$ is the mean atomic weight per free electron.

\begin{figure}
 \includegraphics[width=\columnwidth]{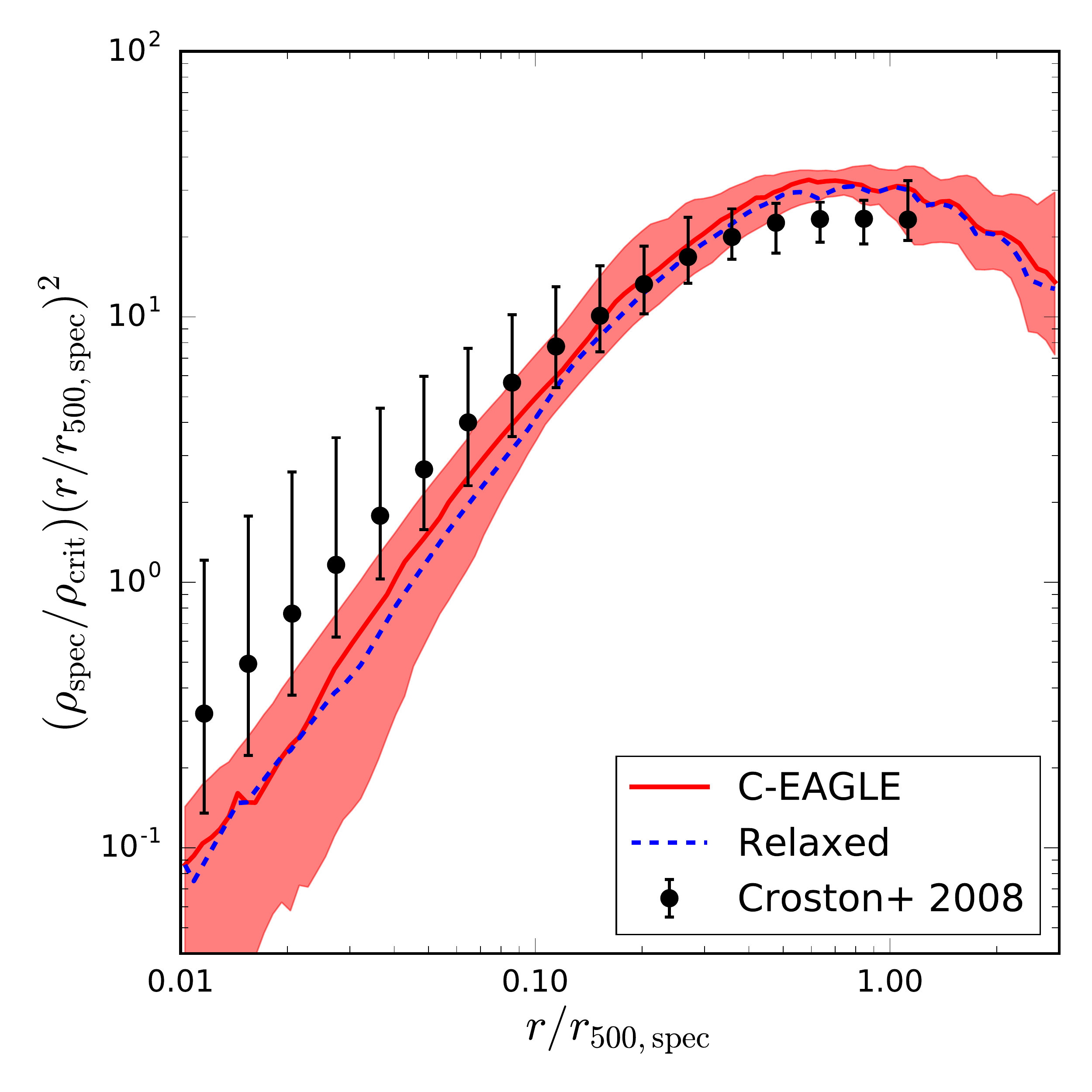}
 \caption{Median density profiles for the \textsc{c-eagle} sample (red solid) and the subset of clusters defined as relaxed (blue dashed) at $z=0.1$, scaled by $(r/r_{500,\rm{spec}})^2$ to reduce the dynamic range. The red shaded region shows the $16^{\mathrm{th}}$ and $84^{\mathrm{th}}$ percentiles of the full sample. The black circles show the median observed profile from the REXCESS sample \citep{Croston2008}, with the error bars enclosing $1\sigma$ of the sample.}
 \label{fig:pro_rho}
\end{figure}

In Fig. \ref{fig:pro_rho} we plot the three-dimensional dimensionless median density profile at $z=0.1$ for all \textsc{c-eagle} clusters and the relaxed subsample. The profiles are scaled by $(r/r_{500,\mathrm{spec}})^{2}$ to reduce the dynamic range. At $r <0.1\,r_{500,\mathrm{spec}}$ the density is somewhat low compared to the observations, but at radii $>0.4\,r_{500,\mathrm{spec}}$ the hot gas is slightly denser than observed. We also find no significant difference between the relaxed and full samples. The lower than observed density in the cluster cores suggests that the AGN feedback is injecting enough energy to displace and heat central gas and that the injection rate is sufficient to prevent gas from cooling and flowing inwards. However, in section \ref{sec:screlations} we found that the \textsc{c-eagle} clusters are too gas rich. The excess gas is at radii $>0.5r_{500,\mathrm{spec}}$ where the \textsc{c-eagle} clusters have a density that exceeds the observed density profile. These results suggest that the AGN feedback in the \textsc{eagle} model is capable of ejecting gas from the core of the cluster, but is not moving sufficient amounts beyond $r_{500}$.

\begin{figure}
 \includegraphics[width=\columnwidth]{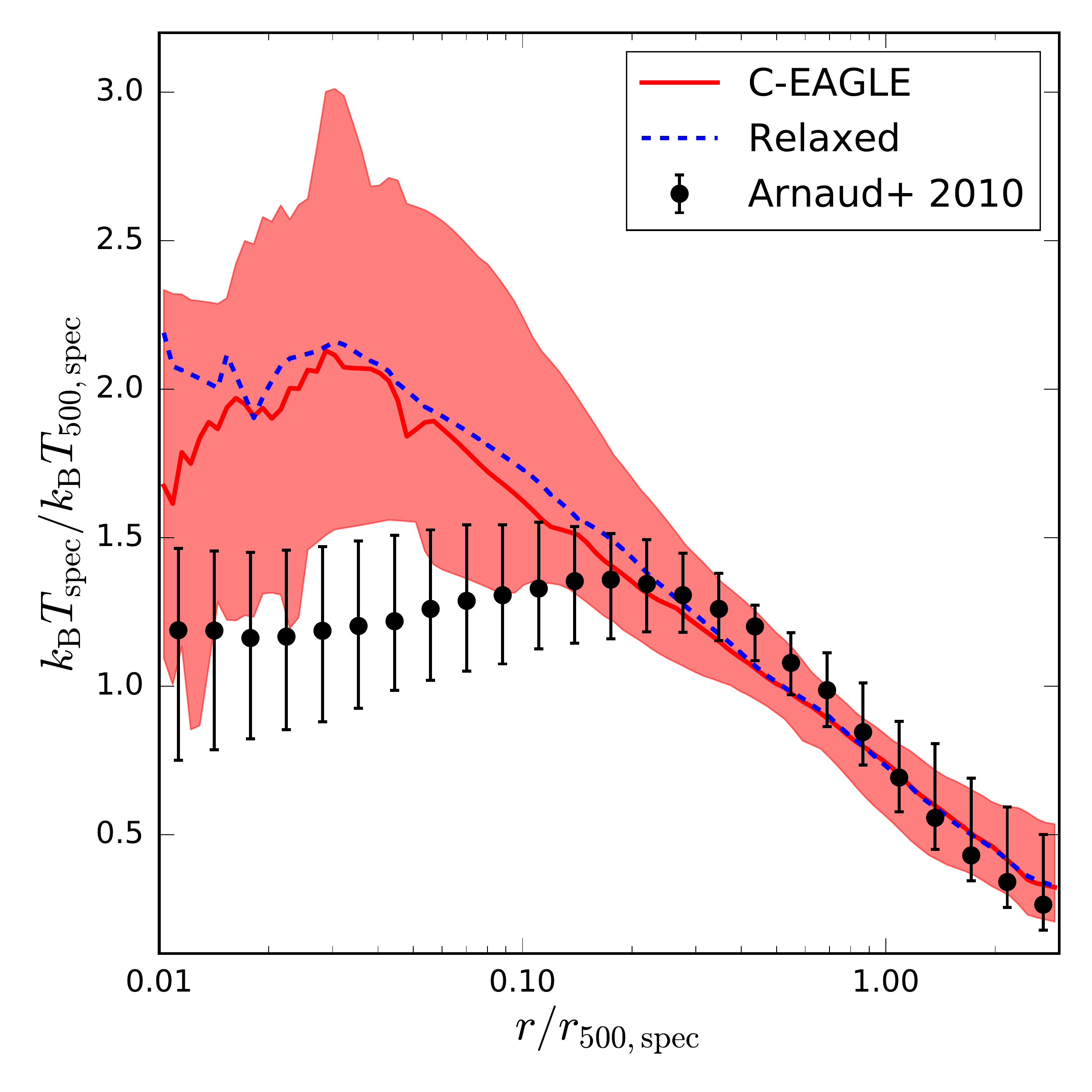}
 \caption{Median spectroscopic temperature profiles for the \textsc{c-eagle} sample and the relaxed subset at $z=0.1$. Marker styles are identical to Fig. \ref{fig:pro_rho}. The median observed profile is calculated by combining the pressure profiles from \citet{Arnaud2010} with the entropy profiles from \citet{Pratt2010}, with the error bars enclosing $1\sigma$ of the sample.}
 \label{fig:pro_temp}
\end{figure}

We plot the median dimensionless spectroscopic temperature profile at $z=0.1$ for the \textsc{c-eagle} clusters and the relaxed subset in Fig. \ref{fig:pro_temp}. Again, we find negligible difference between the full sample and the relaxed sample. At radii greater than $0.2\,r_{500,\mathrm{spec}}$ we find good agreement between the observed median profile and the \textsc{c-eagle} clusters, with both profiles showing a similar level of scatter. However, inside this radius the two profiles diverge significantly from each other. The observed clusters turn over at $0.2\,r_{500,\mathrm{spec}}$, with a roughly flat profile in the cluster core. On the other hand, the \textsc{c-eagle} clusters have a median profile that continues to rise until $0.03\,r_{500,\mathrm{spec}}$ and reaching a peak value $\approx60$ per cent higher than the observed median profile. The temperature profile is consistent with the density profile, AGN feedback is ejecting and heating central gas in the cluster cores and producing a temperature profile that rises all the way to the centre of the cluster.

\begin{figure}
 \includegraphics[width=\columnwidth]{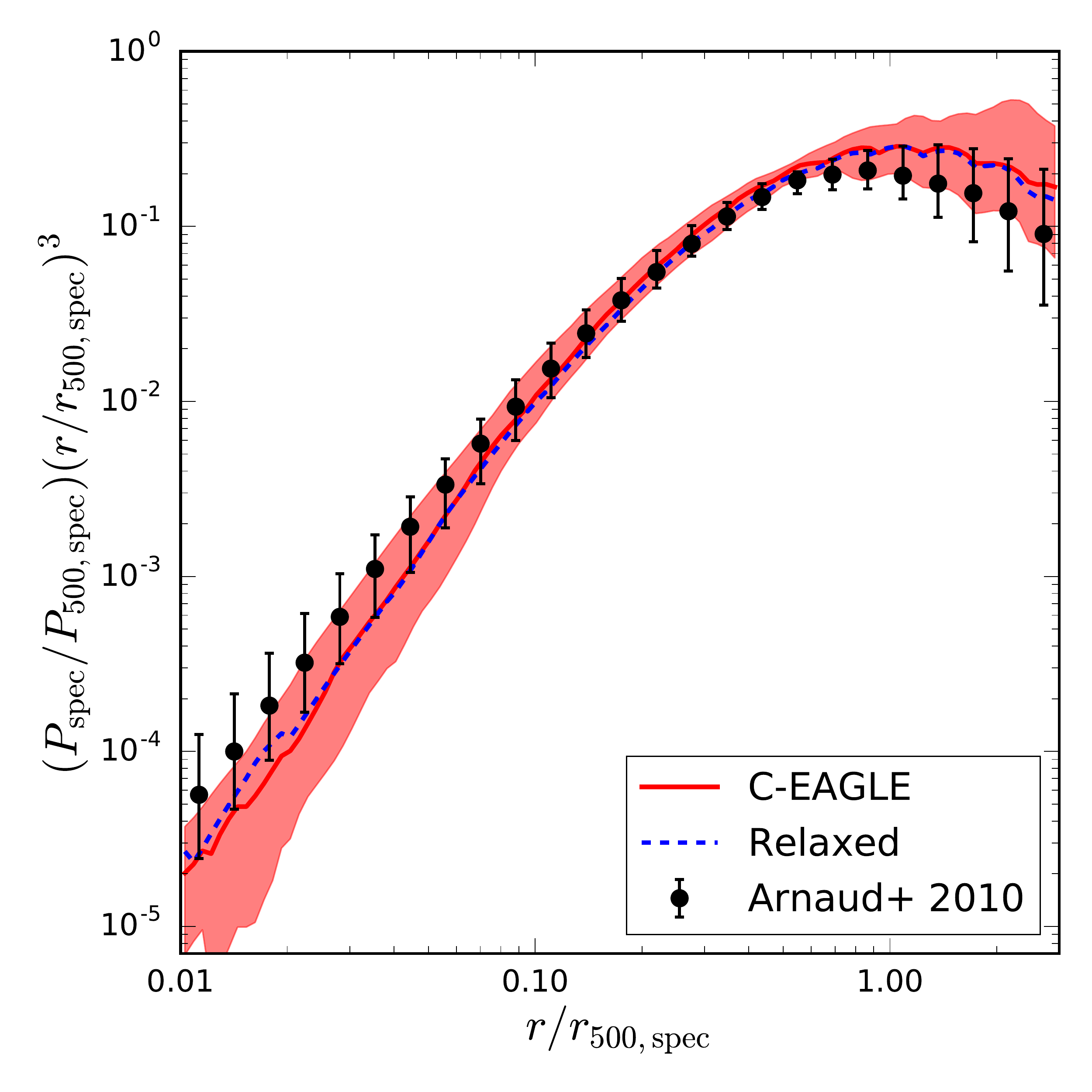}
 \caption{Median pressure profiles for the \textsc{c-eagle} sample and the relaxed subset at $z=0.1$. Marker styles are identical to Fig. \ref{fig:pro_rho}. Data points show the median observed profile from \citet{Arnaud2010}, with the error bars enclosing $1\sigma$ of the sample.}
 \label{fig:pro_pres}
\end{figure}

In Fig. \ref{fig:pro_pres} we plot the median dimensionless pressure profiles, scaled by $(r/r_{500,\mathrm{spec}})^{3}$, at $z=0.1$ for the \textsc{c-eagle} clusters and the relaxed subset. We find reasonable agreement between the pressure profile of the simulated clusters and the observed pressure profile, with the median profile within the intrinsic scatter of the observed sample at all radii. In the core of the cluster the profiles begin to diverge, likely due to the injection of energy by the AGN once the cluster core has formed. The agreement of the pressure profiles with the observations, despite clear differences in both the density and temperature profiles, shows that the clusters are approximately in pressure equilibrium. This suggests that central temperature profile is being driven away from the observed profile by constant heating from AGN feedback, rather than capturing the clusters after one large feedback event.

\begin{figure}
 \includegraphics[width=\columnwidth]{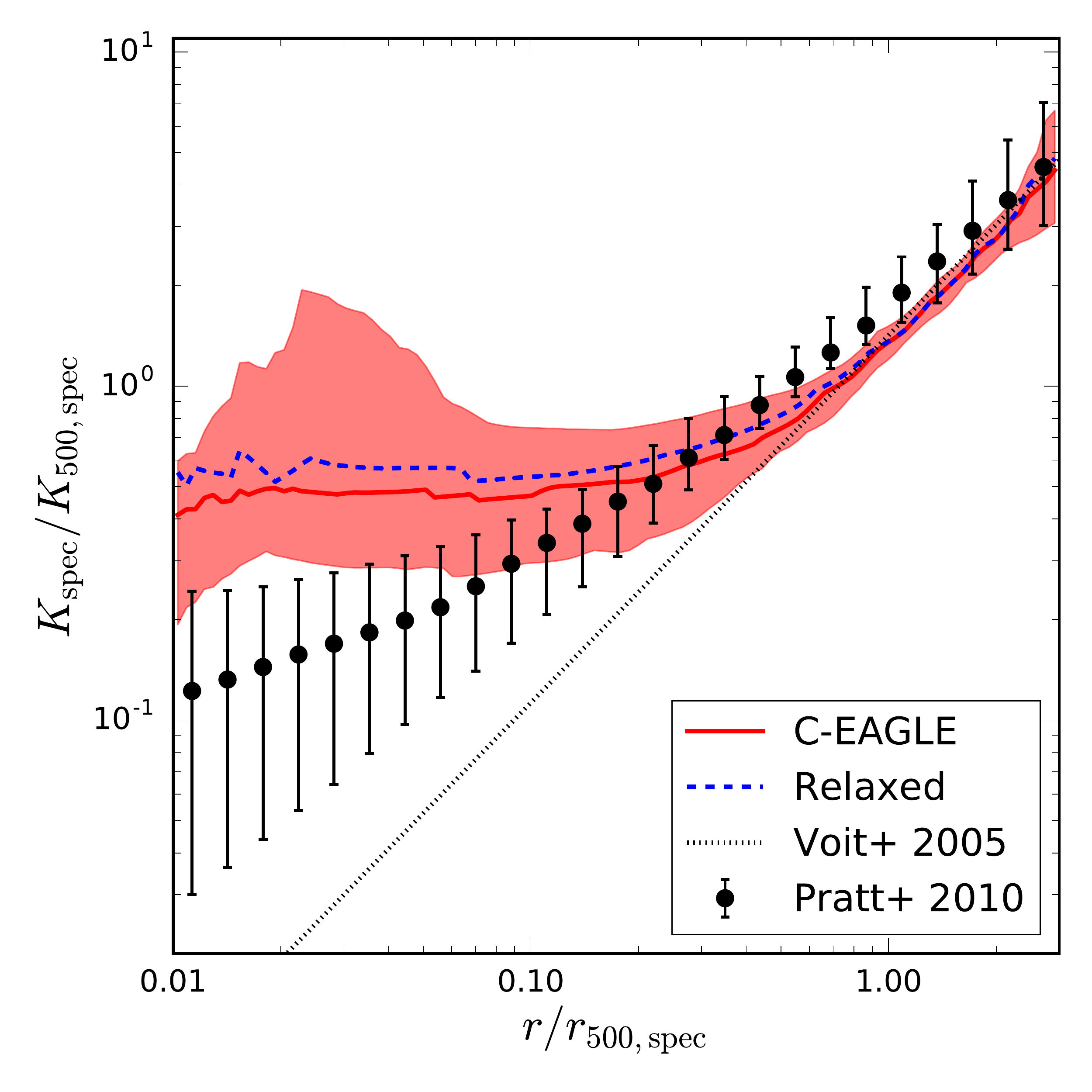}
 \caption{Median entropy profiles for the \textsc{c-eagle} sample and its relaxed subset at $z=0.1$. Marker styles are identical to Fig. \ref{fig:pro_rho}. Data points show the median observed profile from \citet{Pratt2010}, with the error bars enclosing $1\sigma$ of the sample. Additionally, we show the prediction from non-radiative simulations \citep{VoitKayBryan2005}.}
 \label{fig:pro_enty}
\end{figure}

We plot the median $z=0.1$ entropy profiles of the \textsc{c-eagle} clusters and the relaxed subset in Fig. \ref{fig:pro_enty}. Beyond a radius of $0.5\,r_{500,\mathrm{spec}}$ the \textsc{c-eagle} clusters tend to the power-law result predicted by non-radiative simulations \citep{VoitKayBryan2005}. However, inside this radius the entropy profile develops a large core. The radial extent of the core region is significantly greater than observed profile and central value of the entropy is up to a factor of 5 larger. Observational results have shown that the central cooling times of clusters take a range of values between those that are significantly shorter than the Hubble time, `cool-core' clusters with power-law-like entropy profiles, and those that are longer than the Hubble time, `non-cool-core' clusters with cored entropy profiles \citep[e.g.][]{White1997,Peres1998,Sanderson2009,Cavagnolo2009}. Previous simulation work has reproduced this range of cooling times \citep{Rasia2015,Hahn2017}. All of the \textsc{c-eagle} clusters are non-cool-core clusters with cored entropy profiles. In agreement with the temperature profiles, the entropy profiles suggests that the AGN feedback is heating the core of the cluster. Several of the clusters have inverted central entropy profiles, which can only be produced by a recent injection of energy as it is thermodynamically unstable. This inject of energy by the AGN quickly destroys any cool-core that begins to form.

\begin{figure}
 \includegraphics[width=\columnwidth]{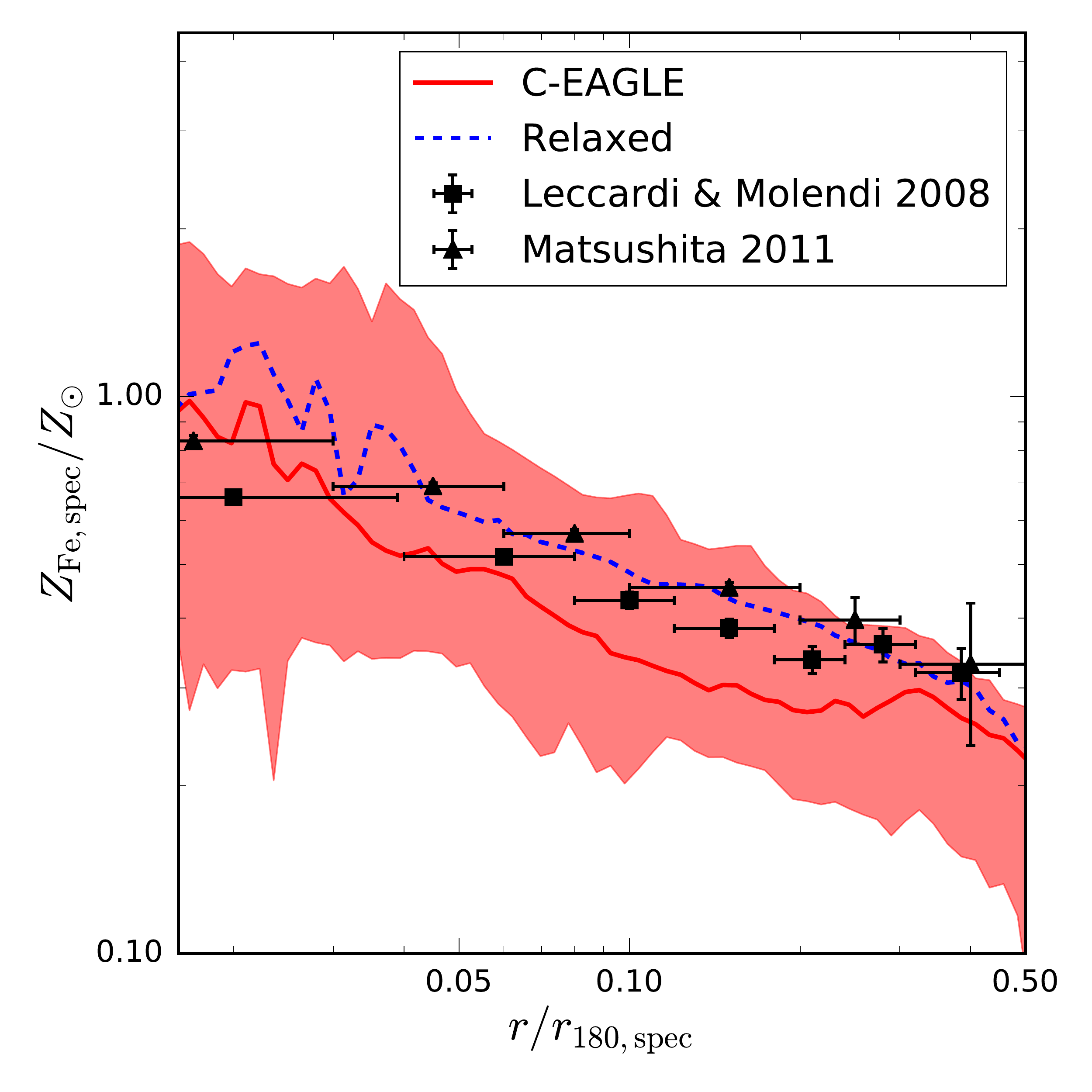}
 \caption{Median emission-weighted iron abundance profiles for the \textsc{c-eagle} sample and its relaxed subset at $z=0.1$. Marker styles are identical to Fig. \ref{fig:pro_rho}. The black squares and triangles are the mean observed profiles from \citet{LeccardiMolendi2008} and \citet{Matsushita2011}, with the error bars showing the measurement uncertainty.}
 \label{fig:pro_zmet}
\end{figure}

Finally, we examine the distribution of metals throughout the cluster volume. In Fig. \ref{fig:pro_zmet} we plot the median emission-weighted iron abundance profiles at $z=0.1$ for the \textsc{c-eagle} clusters and the relaxed subset. We compare with the observations of \citet{LeccardiMolendi2008} who studied $48$ clusters at $z<0.3$ and \citet{Matsushita2011} who studied $28$ clusters at $z<0.08$. We have scaled all results to the solar abundances of \citet{Asplund2009}. For consistency with the observed profiles, we follow \citet{Arnaud2005} and calculate $r_{180,\mathrm{spec}}$ via
\begin{equation}
 r_{180,\mathrm{spec}} = 1.78\left(\frac{k_{\mathrm{B}}T^{\mathrm{X,ce}}_{500,\mathrm{spec}}}{5\,\mathrm{keV}}\right)^{0.5}E^{-1}(z)\,,
\end{equation}
where $k_{\mathrm{B}}T^{\mathrm{X,ce}}_{500,\mathrm{spec}}$ is the core-excised spectroscopic X-ray temperature measured by a single temperature fit to the combined spectrum of all particles that fall within $(0.15-1.0)\,r_{500,\mathrm{spec}}$.

We find reasonable agreement between the observed average metallicity profiles and the median profile of the \textsc{c-eagle} sample, though the observations appear to have a marginally flatter profile. The simulated profiles show a similar level of scatter to the observed scatter \citep[see Fig. 2 of][]{LeccardiMolendi2008}. The relaxed subset shows a similar trend to the full sample, but with a normalization that is systematically higher by $0.1$ dex, however the difference is within the intrinsic scatter of the simulated profiles.

\section{Discussion}
\label{sec:Dis}
We have presented the global properties of the \textsc{c-eagle} clusters, a set of zoom simulations of $30$ objects run with the \textsc{eagle} galaxy formation model that yields realistic galaxies and resolves the ICM on kiloparsec scales. We did not recalibrate the model on cluster scales; instead we chose to retain \textsc{eagle}'s good match to the observed galaxy stellar mass function, the galaxy size-mass relation and the black hole mass-stellar mass relation, thus extending the \textsc{eagle} results to higher mass haloes. We found a reasonable match to observations for the total stellar content of the clusters and black hole mass-stellar mass relation. The ICM properties, a prediction of a calibrated galaxy formation model, are a reasonable match to the properties of observed low-redshift clusters. The main exceptions are the gas fractions and the cluster entropy profiles. Both of these results suggest that the AGN feedback needs revision if it is to reproduce more realistic cluster properties.

The normalization of the gas fraction-total mass relation is high compared to the observed relation (see Fig. \ref{fig:MgM} and \ref{fig:MgM_true}). Previous work by \citet{McCarthy2011} has shown that the majority of gas expulsion from galaxy groups and clusters occurs between $z\approx2-4$. In our model, the first AGN feedback events occur when the black hole enters its rapid growth phase, which occurs when the host halo becomes massive enough that its halo gas is too hot for supernova feedback to regulate the inflow \citep{Bower2017}. Since the progenitors of higher mass haloes collapse at an earlier epoch than those of smaller haloes, they cross the mass threshold for rapid BH growth earlier when the density of the Universe is higher. This means that their potential wells are deeper, requiring more energetic feedback to expel the same fraction of gas. The higher-than-observed gas fractions of the \textsc{c-eagle} clusters suggest that AGN feedback does not inject sufficient energy into cluster progenitors at high redshift. This may also explain why the BCGs are too massive (Bah\'e et al. 2017), since more efficient feedback at higher redshift will also suppress star formation in cluster progenitors.

In addition, a comparison of the simulated and observed entropy profiles (Fig. \ref{fig:pro_enty}) shows that the simulated clusters have significantly larger cores, higher central entropies than observed and none of them are cool cores. This suggests that AGN feedback continues injecting energy into the cluster cores once they have formed, maintaining or increasing their size and preventing cool cores from ever reforming. AGN feedback in \textsc{eagle} uses a modified \citet{BoothSchaye2009} model, which is based on \citet{SpringelDiMatteoHernquist2005}. These two models have been used by several previous studies \citep{Planelles2014,Pike2014,LeBrun2014,McCarthy2017} that broadly reproduce the observed entropy profiles of clusters. The principal difference between this study and previous work is that we achieve a mass resolution that is a factor $100-500$ times better. This lowers the energy threshold for an AGN feedback event by the same factor in our implementation. The decrease in energy threshold results in an individual events that are less energetic but occur more frequently, thus rendering the feedback smoother and less episodic. In addition, the high resolution of the \textsc{c-eagle} simulations means that the gas reaches higher densities around the central BH compared to previous studies. In a Bondi-Hoyle model, the rate of accretion scales as the mass of the BH squared. Therefore, once the cluster core has formed and the central BH is massive, even modest larger scale densities will be enhanced in the immediate vicinity of the BH and so lead to significant accretion. Thus it is not necessary for cool core to reform in order to trigger an AGN feedback event, as is the case at lower resolution.

If the AGN feedback events were more energetic, they might expel gas more effectively from high-redshift progenitors, potentially making gas infall and star formation less efficient. Increasing the energy threshold for individual events would also make feedback more intermittent, perhaps allowing cool cores to reform during periods of low activity. However, in the current \textsc{eagle} model we are limited by the pressure-entropy SPH formalism, which fails to conserve energy if too many particles are heated in a single time step. In addition there are other effects to consider. The AGN feedback model currently distributes heat isotropically to the surrounding gas. If the heating was concentrated into a jet, it might allow a cool core to coexist with the transport of energy to larger radii as is observed, for example, in the Perseus cluster \citep[e.g.][]{Zhuravleva2014}. The role of our artificial conduction scheme in shaping the cluster core is also unclear. Improvement of the \textsc{eagle} AGN feedback model is ongoing, but this work highlights that we must continue to improve the modelling of processes that occur below the resolution limit of the simulations as we push to higher resolution.

\section{Summary}
\label{sec:Sum}
In this paper we have introduced the Cluster EAGLE (\textsc{c-eagle}) simulation project, a set of zoom simulations of the formation of $30$ galaxy clusters in the mass range $10^{14}<M_{200}/\mathrm{M}_{\astrosun}<10^{15.4}$ that resolve the cluster galaxies and the ICM on kiloparsec scales using the state-of-the-art \textsc{eagle} AGNdT9 model of \citetalias{Schaye2015}. In this paper we have presented the global cluster properties, a prediction of a model calibrated for the formation of field galaxies. The properties of the cluster galaxies are analysed in Bah\'e et al. (2017). Our main results are:
\begin{enumerate}
 \item The \textsc{c-eagle} clusters provide a consistent extension to periodic volumes of the original \textsc{eagle} project, sampling the massive objects that were missing due to their limited volumes.
 \item Estimating masses from mock X-ray observations rather than using the true mass results in a hydrostatic bias of $15-20\%$ (Fig. \ref{fig:Mbias}). Selecting relaxed clusters produces a similar bias but significantly reduces the scatter.
 \item The clusters show reasonable agreement with the observed relations that were used to calibrate the \textsc{eagle} model at lower masses. The clusters show a flat stellar fraction-total mass relation in agreement with the observations (Fig. \ref{fig:MsM}) and the trend, normalization and scatter of the black hole mass-stellar mass relation is in agreement with the observations (Fig. \ref{fig:BHrel}). The \textsc{c-eagle} clusters contain a population of orphan super massive black holes, likely the remnants of galaxy mergers. Several of these orphans reside inside the BCGs.
 \item The \textsc{c-eagle} clusters reproduce the observed gas fraction-total mass trend, but the normalization is too high by $\approx30$ per cent (Fig. \ref{fig:MgM}). Although we selected the AGNdT9 calibration as it better reproduced the gas fractions of low-mass galaxy groups, the different \textsc{eagle} AGN calibrations produce similar results on cluster scales. The progenitors of clusters will form at an earlier epoch than those of less massive haloes, causing them to have deeper potentials due to the increased density of the Universe. The increased gas fractions suggests that in this regime the \textsc{eagle} AGN feedback model does not inject sufficient energy into the progenitors of the \textsc{c-eagle} clusters.
 \item The predicted global X-ray and SZ properties of the clusters (Figs. \ref{fig:TxM}-\ref{fig:YM}) show a good match to the spectroscopic temperature, X-ray luminosity and $Y_{\mathrm{SZ}}$ total mass relations. They reproduce the trend of the $Y_{\mathrm{X}}$-mass relation, but the normalization is marginally high due to the high gas fractions.
 \item The \textsc{c-eagle} clusters show good agreement with observed metallicities (Fig. \ref{fig:ZxTx}) and their distribution throughout the cluster volume (Fig. \ref{fig:pro_zmet}).
 \item Median hot gas profiles show that our cluster cores are uniformly less dense and hotter than observed (Figs. \ref{fig:pro_rho}-\ref{fig:pro_enty}). As a result, our clusters have larger than observed entropy cores and our sample includes no cool core clusters. This suggests that the AGN feedback is insufficiently episodic, injecting energy continually into hot gas cores once they have formed. The median pressure profile is in reasonable agreement with the observed profile, showing that the differences between the simulations and the observations are primarily in the thermodynamic rather than in the dynamical state of the gas.
\end{enumerate}
The \textsc{c-eagle} clusters are an attempt to simulate the formation of rich galaxy clusters with a model that yields a galaxy population that is a good match to the observed field population. The resolution required to produce realistic galaxies is more than an order of magnitude higher than that in most previous cluster studies. Overall, the global properties of the ICM, such as the temperature, luminosity, and the total content and distribution of metals, are in reasonable agreement with the observations. A companion paper \citep{Bahe2017} demonstrates that the galaxy properties are also in reasonable agreement with observations, although our BCGs are clearly too massive. Future \textsc{c-eagle} papers will address the formation and evolution of cluster galaxies, the interaction of galaxies with the ICM, the formation of structures in the ICM, and mass estimation systematics for cluster cosmology. We will also investigate the effects of the AGN feedback in more detail, with the aim of producing an improved model that reproduces even more closely the observed properties of cluster galaxies and the ICM.

\section*{Acknowledgements}
This work used the DiRAC Data Centric system at Durham University, operated by the Institute for Computational Cosmology on behalf of the STFC DiRAC HPC Facility (www.dirac.ac.uk). This equipment was funded by BIS National E-infrastructure capital grant ST/K00042X/1, STFC capital grants ST/H008519/1 and ST/K00087X/1, STFC DiRAC Operations grant ST/K003267/1 and Durham University. DiRAC is part of the National E-Infrastructure. The Hydrangea simulations were in part performed on the German federal maximum performance computer `HazelHen' at the maximum performance computing centre Stuttgart (HLRS), under project GCS-HYDA / ID 44067 financed through the large-scale project `Hydrangea' of the Gauss Center for Supercomputing. Further simulations were performed at the Max Planck Computing and Data Facility in Garching, Germany. We also gratefully acknowledge PRACE for awarding the EAGLE project access to the Curie facility based in France at Tr\`es Grand Centre de Calcul. DJB and STK acknowledge support from STFC through grant ST/L000768/1. CDV acknowledges financial support from the Spanish Ministry of Economy and Competitiveness (MINECO) under the 2011 and 2015 Severo Ochoa Programs SEV-2011-0187 and SEV-2015-0548, and grants AYA2014-58308 and RYC-2015-18078. PAT (ORCID 0000-0001-6888-6483) acknowledges support from the Science and Technology Facilities Council (grant number ST/L000652/1). RAC is a Royal Society University Research Fellow. The study was sponsored by the Netherlands Organisation for Scientific Research (NWO), through VICI grant 639.043.409, and the European Research Council under the European Union's Seventh Framework Programme (FP7/2007- 2013) / ERC Grant agreement 278594-GasAroundGalaxies. Support was also received via the Interuniversity Attraction Poles Programme initiated by the Belgian Science Policy Office ([AP P7/08 CHARM]), the National Science Foundation under Grant No. NSF PHY11-25915, and the UK Science and Technology Facilities Council (grant numbers ST/F001166/1 and ST/I000976/1) via rolling and consolidated grants awarded to the ICC.

\bibliographystyle{mnras}
\bibliography{ms}

\appendix
\section{Table of observable quantities}
\label{app:table}
The \textsc{c-eagle} sample was selected from a large dark matter only simulation and resimulated using the zoom technique. To enable a fair comparison of the simulations with observational data, we compute a mock X-ray spectrum for every particle in the simulation. These are then summed to provide a mock spectrum for the cluster. We then perform a hydrostatic analysis using the density and temperature profiles measured from the mock spectra to produced estimated masses and radii. Observable quantities are then calculated within estimated apertures, see \citet{LeBrun2014} for further details. The tables below provide the locations of the sample in the parent simulation $z=0$, the size of the sphere selected at $z=0$ from the parent to produce the high-resolution region and the global properties of the clusters at $z=0.1$ that are presented in Section \ref{sec:screlations}. For additional cluster properties we direct the reader to Appendix A of \citep{Bahe2017}.

\begin{table*}
 \caption{Global quantities of the \textsc{c-eagle} clusters at $z=0.1$, with `hse' and `spec' quantities estimated via the mock X-ray pipeline. The IC location and and IC extent specify the comoving centre and radius of the sphere that contains the high-resolution Lagrangian region in the initial conditions. The \textsc{panphasia} phase descriptor is [Panph1, L14, (2152, 5744, 757), S3, CH1814785143, EAGLE\_L3200\_VOL1].}
 \centering
 \begin{tabularx}{\textwidth}{l C C C C C C C C C}
 \hline
 Cluster & $M_{500,\mathrm{true}}$ & $M_{500,\mathrm{hse}}$ & $M_{500,\mathrm{spec}}$ & $r_{500,\mathrm{true}}$ & $r_{500,\mathrm{spec}}$ & \multicolumn{3}{c}{IC location} & IC extent \\
  & \multicolumn{3}{c}{$[\log_{10}(M/\mathrm{M}_{\astrosun})]$} & \multicolumn{2}{c}{$[\mathrm{Mpc}]$} & $x$ $[\mathrm{Mpc}/h]$ & $y$ $[\mathrm{Mpc}/h]$ & $z$ $[\mathrm{Mpc}/h]$ & $[\mathrm{Mpc}/h]$ \\
 \hline
 CE-00 & 13.905 & 13.756 & 13.751 & 0.65 & 0.58 & ~207.81 & 1498.48 & 1793.66 & 22.45 \\
 CE-01 & 13.982 & 13.927 & 13.931 & 0.69 & 0.66 & 1765.99 & 1721.52 & 1541.50 & 21.03 \\
 CE-02 & 13.920 & 13.870 & 13.871 & 0.66 & 0.63 & 1962.55 & 1953.81 & ~234.28 & 21.03 \\
 CE-03 & 13.926 & 13.877 & 13.861 & 0.66 & 0.63 & 1772.74 & 1915.43 & ~616.81 & 27.33 \\
 CE-04 & 13.853 & 13.723 & 13.697 & 0.62 & 0.55 & 1170.49 & 1524.83 & 1807.16 & 23.97 \\
 CE-05 & 13.892 & 13.902 & 13.880 & 0.64 & 0.64 & ~395.90 & ~613.11 & 1126.81 & 23.97 \\
 CE-06 & 14.128 & 14.102 & 14.083 & 0.77 & 0.74 & 1774.60 & 1519.13 & ~206.88 & 35.53 \\
 CE-07 & 14.176 & 14.127 & 14.121 & 0.80 & 0.77 & ~856.66 & 1662.35 & ~869.10 & 31.16 \\
 CE-08 & 14.070 & 13.980 & 13.941 & 0.74 & 0.67 & ~329.85 & ~497.42 & ~247.39 & 24.63 \\
 CE-09 & 14.244 & 14.118 & 14.084 & 0.84 & 0.74 & ~922.84 & ~982.93 & 1499.01 & 24.77 \\
 CE-10 & 14.301 & 14.250 & 14.225 & 0.88 & 0.83 & 1771.02 & 1093.98 & 1271.48 & 25.60 \\
 CE-11 & 14.290 & 14.182 & 14.182 & 0.87 & 0.80 & 1740.71 & ~459.04 & ~920.32 & 26.45 \\
 CE-12 & 14.422 & 14.349 & 14.326 & 0.96 & 0.90 & ~793.25 & ~945.74 & ~682.99 & 36.72 \\
 CE-13 & 14.341 & 14.249 & 14.266 & 0.91 & 0.86 & ~682.60 & 1023.43 & 1327.07 & 32.20 \\
 CE-14 & 14.563 & 14.571 & 14.568 & 1.08 & 1.08 & ~184.91 & ~991.14 & 1381.74 & 49.32 \\
 CE-15 & 14.407 & 14.404 & 14.418 & 0.95 & 0.96 & 1364.14 & ~501.66 & 1179.09 & 41.86 \\
 CE-16 & 14.311 & 14.170 & 14.143 & 0.89 & 0.78 & ~487.63 & 1526.81 & ~406.49 & 41.86 \\
 CE-17 & 14.537 & 14.461 & 14.433 & 1.05 & 0.97 & ~143.61 & 1251.36 & 1953.15 & 31.16 \\
 CE-18 & 14.639 & 14.587 & 14.555 & 1.14 & 1.07 & ~542.69 & ~580.55 & 1091.34 & 41.86 \\
 CE-19 & 14.586 & 14.483 & 14.445 & 1.09 & 0.98 & ~547.19 & ~219.86 & ~760.69 & 37.94 \\
 CE-20 & 14.482 & 14.361 & 14.342 & 1.01 & 0.91 & 1827.41 & 1204.77 & 2009.96 & 30.16 \\
 CE-21 & 14.800 & 15.012 & 14.934 & 1.29 & 1.43 & ~767.31 & ~607.95 & ~650.43 & 40.51 \\
 CE-22 & 14.837 & 14.721 & 14.701 & 1.33 & 1.20 & 1407.02 & 1572.74 & ~567.84 & 62.05 \\
 CE-23 & 14.426 & 14.291 & 14.256 & 0.97 & 0.85 & 1378.96 & 2033.10 & 1838.00 & 37.94 \\
 CE-24 & 14.821 & 14.728 & 14.666 & 1.31 & 1.16 & ~209.13 & ~668.70 & 1948.65 & 52.66 \\
 CE-25 & 15.045 & 15.070 & 15.068 & 1.56 & 1.58 & ~697.69 & ~861.16 & ~860.76 & 49.32 \\
 CE-26 & 14.899 & 14.838 & 14.780 & 1.39 & 1.27 & 1907.88 & ~852.55 & 1346.40 & 43.26 \\
 CE-27 & 14.689 & 14.496 & 14.389 & 1.18 & 0.94 & 1790.48 & ~618.80 & 1783.99 & 41.86 \\
 CE-28 & 14.902 & 14.688 & 14.671 & 1.39 & 1.17 & ~944.94 & ~707.08 & 1388.09 & 62.05 \\
 CE-29 & 15.077 & 15.089 & 14.912 & 1.60 & 1.40 & ~719.79 & 1449.38 & 1015.75 & 60.04 \\
 \hline
 \end{tabularx}
 \label{tab:Global_properties_I}
\end{table*}

\renewcommand\arraystretch{1.2}
\begin{table*}
 \caption{Global quantities of the \textsc{c-eagle} clusters at $z=0.1$, estimated from the mock X-ray pipeline. All quantities are measured inside $r_{500,\mathrm{spec}}$ except for $Y_{\mathrm{SZ}}$, which is measured inside $5r_{500,\mathrm{spec}}$.}
 \centering
 \begin{tabularx}{\textwidth}{l C C C C C C C P{1.0cm}}
 \hline
 Cluster & $k_{\mathrm{B}}T^{\mathrm{X}}$ & $L^{0.5-2.0\mathrm{keV}}_{\mathrm{X}}$ & $M_{\mathrm{gas}}$ & $M_{\mathrm{star}}$ & $Y_{\mathrm{X}}$ & $Y_{\mathrm{SZ}}$ & $Z_{\mathrm{Fe}}$ & $E_{\mathrm{kin}}/E_{\mathrm{thrm}}$ \\
  & $[\mathrm{keV}]$ & $[\log_{10}(L/\mathrm{ergs}^{-1})]$ & $[\log_{10}(M/\mathrm{M}_{\astrosun})]$ & $[\log_{10}(M/\mathrm{M}_{\astrosun})]$ & $[\log_{10}(Y/\mathrm{M}_{\astrosun}\mathrm{keV})]$ & $[\log_{10}(Y/\mathrm{Mpc}^2)]$ & $[\mathrm{Z}_{\mathrm{Fe},\astrosun}]$ &  \\
 \hline
 CE-00 & 1.81 & 43.234 & 12.881 & 12.230 & 13.125 & -5.445 & 0.26 & 0.11 \\
 CE-01 & 2.18 & 43.459 & 13.026 & 12.290 & 13.302 & -5.521 & 0.21 & 0.24 \\
 CE-02 & 1.90 & 43.227 & 12.954 & 12.164 & 13.217 & -5.425 & 0.27 & 0.05 \\
 CE-03 & 2.01 & 43.114 & 12.924 & 12.140 & 13.221 & -5.298 & 0.21 & 0.06 \\
 CE-04 & 1.79 & 42.888 & 12.789 & 11.994 & 13.027 & -5.449 & 0.22 & 0.11 \\
 CE-05 & 1.75 & 43.240 & 12.970 & 12.154 & 13.213 & -5.340 & 0.29 & 0.09 \\
 CE-06 & 2.28 & 43.518 & 13.204 & 12.306 & 13.551 & -5.046 & 0.22 & 0.09 \\
 CE-07 & 2.48 & 43.657 & 13.247 & 12.354 & 13.631 & -4.984 & 0.21 & 0.09 \\
 CE-08 & 2.03 & 43.518 & 13.120 & 12.247 & 13.412 & -5.126 & 0.25 & 0.09 \\
 CE-09 & 2.60 & 43.620 & 13.237 & 12.434 & 13.641 & -4.850 & 0.20 & 0.16 \\
 CE-10 & 3.00 & 43.890 & 13.414 & 12.479 & 13.886 & -4.758 & 0.26 & 0.05 \\
 CE-11 & 2.66 & 44.080 & 13.389 & 12.482 & 13.815 & -4.845 & 0.29 & 0.21 \\
 CE-12 & 3.28 & 43.942 & 13.485 & 12.644 & 13.990 & -4.636 & 0.26 & 0.08 \\
 CE-13 & 3.20 & 44.117 & 13.443 & 12.460 & 13.917 & -4.721 & 0.29 & 0.06 \\
 CE-14 & 3.99 & 44.321 & 13.704 & 12.709 & 14.302 & -4.508 & 0.23 & 0.31 \\
 CE-15 & 3.56 & 44.037 & 13.511 & 12.592 & 13.993 & -4.608 & 0.16 & 0.22 \\
 CE-16 & 3.06 & 43.877 & 13.351 & 12.505 & 13.814 & -4.522 & 0.26 & 0.12 \\
 CE-17 & 3.91 & 44.284 & 13.623 & 12.684 & 14.208 & -4.525 & 0.21 & 0.25 \\
 CE-18 & 4.22 & 44.402 & 13.744 & 12.787 & 14.359 & -4.262 & 0.19 & 0.09 \\
 CE-19 & 3.23 & 44.102 & 13.660 & 12.723 & 14.171 & -4.354 & 0.25 & 0.30 \\
 CE-20 & 3.60 & 44.127 & 13.551 & 12.660 & 14.077 & -4.457 & 0.24 & 0.12 \\
 CE-21 & 5.09 & 44.540 & 13.990 & 13.024 & 14.694 & -4.009 & 0.51 & 0.29 \\
 CE-22 & 6.04 & 44.618 & 13.937 & 12.967 & 14.668 & -3.854 & 0.16 & 0.12 \\
 CE-23 & 3.16 & 43.904 & 13.452 & 12.642 & 13.934 & -4.325 & 0.19 & 0.26 \\
 CE-24 & 5.31 & 44.412 & 13.876 & 12.937 & 14.598 & -4.005 & 0.17 & 0.13 \\
 CE-25 & 7.95 & 45.021 & 14.214 & 13.177 & 15.102 & -3.650 & 0.23 & 0.31 \\
 CE-26 & 6.43 & 44.589 & 14.010 & 13.037 & 14.809 & -3.765 & 0.19 & 0.08 \\
 CE-27 & 5.28 & 42.987 & 13.094 & 12.323 & 13.811 & -4.259 & 0.19 & 0.24 \\
 CE-28 & 6.29 & 44.407 & 13.906 & 13.051 & 14.697 & -3.735 & 0.18 & 0.13 \\
 CE-29 & 7.66 & 44.942 & 14.188 & 13.185 & 15.067 & -3.510 & 0.52 & 0.30 \\
 \hline
 \end{tabularx}
 \label{tab:global_properties_II}
\end{table*}
\renewcommand\arraystretch{1.0}

\section{Power law fits}
\label{app:plf}
In this work we have presented how observable quantities scale with estimated mass. To get a quantitative measure of the scatter of the \textsc{c-eagle} clusters about each scaling relation we assume that the cluster scaling relations are simple power-laws \citep{Kaiser1986}. At $z=0.1$, we fit each scaling relation with a model of the form
\begin{equation}
 Y = 10^{A}\left(\frac{X}{X_{\mathrm{piv}}}\right)^{\alpha}\,,
\end{equation}
where $Y$ is the observable quantity, $X$ is the total mass, $A$ and $\alpha$ are the normalization and slope of the power-law and $X_{\mathrm{piv}}=4.0\times10^{14}\,\mathrm{M}_{\astrosun}$ is the pivot mass. The scatter about the power-law relation is then calculated via
\begin{equation}
 \sigma_{\log_{10}} = \sqrt{\frac{1}{N}\sum_{i=1}^{N}\left[\log_{10}(Y_{i})-\log_{10}(Y_{\mathrm{mod}})\right]^{2}}\,,
\end{equation}
where $N=30$ is the number of clusters in the sample, $Y_{i}$ is the measured observable quantity, $Y_{\mathrm{mod}}$ is the expected observable value of a cluster of mass $X_{i}$, and $\sigma_{\ln}=\ln(10)\sigma_{\log_{10}}$. We fit power laws to the full sample, those clusters defined as relaxed and those defined as unrelaxed. Table \ref{tab:plf} summarizes the power-law best-fit relations and the scatter about them.

\renewcommand\arraystretch{1.5}
\begin{table*}
 \caption{Summary of the best-fit values of the linear power-law fits and the scatter about each fit for the scaling relations presented in this work. All fits are done at $z=0.1$. The values of $A$ given in the table assume that the units of $10^{A}$ are $\mathrm{keV}$, $\mathrm{ergs}^{-1}$, $\mathrm{M}_{\astrosun}$, $\mathrm{M}_{\astrosun}$, $\mathrm{M}_{\astrosun}\mathrm{keV}$ and $\mathrm{Mpc}^{2}$, respectively.}
 \centering
 \begin{tabularx}{\textwidth}{l C C C C C C C C C}
 \hline
 Scaling & \multicolumn{3}{c}{Full sample} &  \multicolumn{3}{c}{Relaxed} & \multicolumn{3}{c}{Unrelaxed} \\
 relation & $A$ & $\alpha$ & $\sigma_{\log_{10}}$ & $A$ & $\alpha$ & $\sigma_{\log_{10}}$ & $A$ & $\alpha$ & $\sigma_{\log_{10}}$ \\
 \hline
 $k_{\mathrm{B}}T^{\mathrm{X}}_{500,\mathrm{spec}}-M_{500,\mathrm{spec}}$ & $~0.67_{-0.02}^{+0.03}$ & $0.47_{-0.02}^{+0.07}$ & 0.06 & $~0.68_{-0.02}^{+0.01}$ & $0.56_{-0.03}^{+0.02}$ & 0.02 & $~0.68_{-0.02}^{+0.02}$ & $0.42_{-0.02}^{+0.09}$ & 0.07 \\
 $L^{0.5-2.0\mathrm{keV}}_{\mathrm{X},500,\mathrm{spec}}-M_{500,\mathrm{spec}}$ & $44.13_{-0.07}^{+0.24}$ & $1.33_{-0.08}^{+0.13}$ & 0.30 & $44.45_{-0.06}^{+0.11}$ & $1.59_{-0.08}^{+0.21}$ & 0.11 & $44.11_{-0.08}^{+0.26}$ & $1.36_{-0.15}^{+0.12}$ & 0.35 \\
 $M_{\mathrm{gas},500,\mathrm{spec}}-M_{500,\mathrm{spec}}$ & $13.68_{-0.03}^{+0.09}$ & $1.07_{-0.05}^{+0.02}$ & 0.13 & $13.82_{-0.01}^{+0.01}$ & $1.15_{-0.02}^{+0.03}$ & 0.03 & $13.68_{-0.03}^{+0.09}$ & $1.09_{-0.08}^{+0.03}$ & 0.15 \\
 $M_{\mathrm{star},500,\mathrm{spec}}-M_{500,\mathrm{spec}}$ & $12.79_{-0.03}^{+0.05}$ & $0.85_{-0.04}^{+0.06}$ & 0.10 & $12.85_{-0.02}^{+0.02}$ & $0.97_{-0.03}^{+0.03}$ & 0.03 & $12.79_{-0.03}^{+0.06}$ & $0.85_{-0.06}^{+0.06}$ & 0.12 \\
 $Y_{\mathrm{X},500,\mathrm{spec}}-M_{500,\mathrm{spec}}$ & $14.39_{-0.02}^{+0.03}$ & $1.57_{-0.07}^{+0.07}$ & 0.10 & $14.48_{-0.01}^{+0.01}$ & $1.69_{-0.02}^{+0.04}$ & 0.04 & $14.39_{-0.02}^{+0.04}$ & $1.57_{-0.10}^{+0.07}$ & 0.12 \\
 $Y_{\mathrm{SZ},5r500,\mathrm{spec}}-M_{500,\mathrm{spec}}$ & $-4.16_{-0.03}^{+0.04}$ & $1.48_{-0.10}^{+0.11}$ & 0.18 & $-4.13_{-0.06}^{+0.01}$ & $1.66_{-0.09}^{+0.03}$ & 0.06 & $-4.16_{-0.03}^{+0.05}$ & $1.46_{-0.17}^{+0.12}$ & 0.21 \\
 \hline
 \end{tabularx}
 \label{tab:plf}
\end{table*}
\renewcommand\arraystretch{1.0}

\bsp	
\label{lastpage}
\end{document}